\documentclass[pra,twocolumn,superscriptaddress,amsmath,amssymb,aps,floatfix]{revtex4-2}
\usepackage{placeins}
\usepackage{graphicx}% Include figure files
\usepackage{dcolumn}% Align table columns on decimal point
\usepackage{bm}% bold math
\usepackage{amsmath,amssymb}
\usepackage{graphics}
\usepackage{amsfonts}
\usepackage{epsfig}
\usepackage{float}
\usepackage{hyperref}
\usepackage{adjustbox}
\usepackage{rotating}
\usepackage{float}
\usepackage{dblfloatfix}
\usepackage{blindtext}

\usepackage{multirow}
\usepackage{graphicx}
\usepackage{dcolumn}
\usepackage{bbold}
\usepackage{bm}
\usepackage{amsmath}
\usepackage{import}
\usepackage{physics}
\usepackage{verbatim}
\usepackage{listings}
\usepackage{xcolor}
\usepackage{xr-hyper}
\usepackage{hyperref}
\usepackage{multirow}
\usepackage{cleveref}
\usepackage{newfloat}
\usepackage{caption}
\usepackage{subcaption}
\usepackage{algorithm}
\usepackage{algpseudocode}
\DeclareFloatingEnvironment[name={Fig S},fileext=lof]{suppfigure}
\begin{document}
\title{Controlling quantum chaos via Parrondo strategies on noisy intermediate-scale quantum hardware}
\author{Aditi Rath}
\email{aditi.rath@niser.ac.in}
\author{Dinesh Kumar Panda}
\email{dineshkumar.quantum@gmail.com}
\author{Colin Benjamin}
\email{colin.nano@gmail.com}
\affiliation{School of Physical Sciences, National Institute of Science Education and Research, Bhubaneswar, Jatni 752050, India}
\affiliation{Homi Bhabha National Institute, Training School Complex, Anushaktinagar, Mumbai
400094, India}
\begin{abstract}
Advancements in Noisy Intermediate-Scale Quantum (NISQ) computing are steadily pushing these systems toward outperforming classical supercomputers on specific, well-defined computational tasks. In this work, we explore and control quantum chaos in NISQ systems using discrete-time quantum walks (DTQW) on cyclic graphs. To efficiently implement quantum walks on NISQ hardware, we employ the quantum Fourier transform (QFT) to diagonalize the conditional shift operator, optimizing circuit depth and fidelity. We experimentally realize the transition from quantum chaos to order via DTQW dynamics on both odd and even cyclic graphs, specifically 3- and 4-cycle graphs, using the counterintuitive Parrondo's paradox strategy across three different NISQ devices. While the 4-cycle graphs exhibit high-fidelity quantum evolution, the 3-cycle implementation shows significant fidelity improvement when augmented with dynamical decoupling pulses. Our results demonstrate a practical approach to probing and harnessing controlled chaotic dynamics on real quantum hardware, laying the groundwork for future quantum algorithms and cryptographic protocols based on quantum walks.
\end{abstract}

\maketitle
\newpage
\twocolumngrid
\section{Introduction}
In this quantum computing era, quantum devices known as NISQ (noisy intermediate-scale quantum) promise better performance than existing classical supercomputers for certain computational tasks~\cite{nisq1,nisq2}.
These NISQ systems are typically based on superconducting qubits~\cite{sup}, trapped ions~\cite{ion}, nuclear spin qubits in silicon~\cite{spin}, or other solid-state platforms~\cite{solid}. However, they are prone to imperfections at various stages, including noisy state preparation, faulty quantum gate operations, and imprecise measurements on all available qubits.
Quantum walks (QW)~\cite{qw}, both in discrete (DTQW) and continuous-time (CTQW) versions, provide a universal platform for quantum computing ~\cite{1,2,3} and quantum simulation~\cite{pan15,pan16,pan17, pan21,pan23,karimi19}, by exploiting quantum interference and coherence. This leads to properties such as the quadratic spread of probability distribution~\cite{pd} and quadratically faster hitting time~\cite{ht}, which form the key elements for obtaining an advantage over classical algorithms. In addition, QWs have found applications in quantum algorithms~\cite{qa1,qa2,qa3}, quantum cryptography~\cite{me-cb2,p4}, and simulating physical~\cite{Phy} and biological systems~\cite{r7,bio}. 

DTQWs are already implemented efficiently using physical systems supporting their conditional walk operation, such as photonic setups~\cite{exp1,exp2,exp3,exp4,exp5}. However, to utilize DTQWs to design quantum algorithms for general-purpose quantum computers and to implement general quantum protocols in circuit models, we need to achieve efficient implementations of DTQWs on NISQ quantum computers for arbitrary graphs. There have been various approaches to implement quantum walks on quantum circuits~\cite{DW, CI, SK, Luca} on even-site cyclic graphs. Ref.~\cite{DW} provides insight into implementing DTQWs on cyclic graphs utilizing the increment-decrement scheme of generalized multi-qubit CNOT gates, while Ref.~\cite{CI} proposes the method of shunt-decomposition on a variety of graphs, and Ref.~\cite{SK} uses the approach of quantum Fourier transform (QFT) to implement DTQW on even cyclic graphs.  

More recently, the authors of Ref.~\cite{Luca} showed that by exploiting the unitarity of QFT and employing adjacent physical qubits, one can implement DTQW on even cycles with improved fidelity and fewer resources compared to Ref.~\cite{SK}. However, a NISQ implementation on general cyclic graphs (with either odd or even number of sites) has been missing. In this manuscript, we show an efficient implementation of DTQW on both odd and even-site cyclic graphs. This work is the first quantum circuit implementation of odd cyclic DTQWs on NISQ quantum devices. Notably, due to larger circuit depth, the 3-cycle implementation is shown to have better fidelity after incorporating dynamical decoupling (DD)~\cite{DD} pulses into the quantum circuits. Further, we realize the transition from chaos to order using Parrondo strategies on NISQ devices, for both even and odd cyclic graphs.

Parrondo's paradox refers to circumstances where a combination of losing strategies can result in a winning outcome~\cite{Parrondo},\cite{Hammer},~\cite{Abbott},\cite{exptparrando}. In the context of QWs, this implies that a combination of chaotic walks can result in a periodic walk, illustrating the emergence of order from chaos~\cite{panda}. In recent studies, Parrondo's strategies have found their applications in classical chaos control theory~\cite{Kang,chaos1} and quantum cryptography~\cite{panda,me-cb2}, where the emergence of order significantly reduces the resources (quantum gates) required in the communication protocol between Alice and Bob. However, there has been no attempt to implement the phenomenon of the emergence of order from chaos in quantum circuits on real quantum hardware (NISQ devices). Herein, we achieve this via utilizing the counterintuitive strategy: Parrondo strategy both in odd and even cycle graphs with and without DD using three different IBM NISQ devices~\cite{ibm}, namely, \texttt{ibm\_sherbrooke},\texttt{ibm\_brisbane}, and \texttt{ibm\_kyiv}.
\\This paper is structured as follows. In section~\ref{2}, we first lay the background on DTQWs on arbitrary cyclic graphs, using Parrondo's strategy to generate order from chaos. In section~\ref{3}, we introduce the methods of quantum circuit implementation of both even and odd cycles, along with their NISQ implementations on \texttt{ibm\_sherbrooke}, to visualize the transition from chaos to order. In the same section, we analyze and address the problems of implementing odd-site (e.g., 3) cycle DTQW with NISQ hardware. We show improvements in the results of the real hardware via the integration of an error suppression tool: \textit{dynamical decoupling } to quantum circuits. In section~\ref{4}, we analyze and compare the circuit implementation of the even (4) cycle DTQW of existing schemes in terms of fidelity and quantum circuit depth. Further, we analyze our results on implementations of DTQWs and transition from chaos to order, on even (4) and odd (3) cyclic graphs, via the fidelity and circuit depths in NISQ quantum computers. We conclude in section~\ref{con}, discuss our future plans, and give a perspective on implementing Parrondo strategies in different contexts via NISQ devices. Additional details on: the algorithm to control chaos via cyclic DTQWs, the transpiled IBM quantum circuits, calculations of circuit depth, results obtained from \texttt{ibm\_brisbane} and \texttt{ibm\_kyiv} quantum computers, calculation of the Lyapunov exponent for 4-cycle DTQWs, results for chaotic sequences of unitary operators, and comparison of probability distributions via Hellinger fidelity and Bhattacharya fidelity are discussed in Appendices~\ref{Algorithm}-\ref{7}.
% providing a general framework to implement arbitrary cyclic DTQWs on NISQ devices and verify the transition of chaos to order via Parrondo's paradox.
%Herein, we combine two chaotic quantum walk-generating coins to yield a periodic quantum walk on both odd and even cycle graphs.

%Odd v choas intro with use (Abhishek, bio, chaos control, cryptography, secure comm).Define Parrondo Besides, to achieve algorithmic advantage of QW, one need to have faithful implementation of QW using NISQ devices/ q-computers, mention: 3 features  of QW allows a feasible implementation. Why NISQ simulation is necessary? 
% 

\section{Theory}\label{2}
\subsection{DTQW on cyclic graphs}\label{2a}
\begin{figure}
     \centering
     \includegraphics[width=0.5\linewidth]{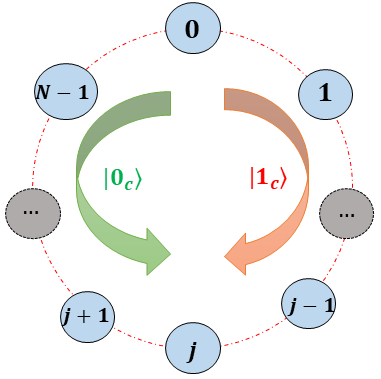}
     \caption{DTQW on an N-cycle graph.}
     \label{cycle}
 \end{figure}

A DTQW on an $N$-cycle (see Fig.~\ref{cycle}) is the tensor product of position ($H_P$) and coin ($H_C$) Hilbert spaces,i.e., $H_P\otimes H_C$. The coin space $H_C$ is defined on the computational basis $\{|s_c\rangle\}$ i.e., $\{|0_c\rangle, |1_c\rangle \} $ and the position space $H_P$ is spanned by basis \(\{|j_p\rangle : j \in \{0,1,2,...,N-1\} \} \). If the walker is initially at the position \( |0_p\rangle \) and in the general superposition of the  coin states, it is represented by the state, \begin{equation}
|\psi(t = 0)\rangle = \cos\left(\frac{\theta}{2}\right) |0_p, 0_c\rangle + e^{i\phi} \sin\left(\frac{\theta}{2}\right) |0_p, 1_c\rangle,
\label{e1}
\end{equation} with \( \theta \in [0, \pi] \) and \( \phi \in [0, 2\pi) \).
 The unitary coin operator is \begin{equation}
C_2(r, a, b) =
\begin{pmatrix}
\sqrt{r} & \sqrt{1-r} e^{i a} \\
\sqrt{1-r} e^{i b} & -\sqrt{r} e^{i(a+b)}
\end{pmatrix}, \label{e2}\end{equation} where \( 0 \leq r \leq 1 \) and \( 0 \leq a,b \leq \pi \).\par 
The walker shifts to the left (right) by one site for coin state \( |0_c\rangle \) (\( |1_c\rangle \)). For the walker on an $N$-cycle graph, the shift operator is \begin{equation} 
    S = \sum_{s=0}^{1} |s\rangle \langle s| \otimes R_s,
\label{e3} \end{equation} where $R_s = \sum_{j=0}^{N-1} |(j+2s-1) \mod N\rangle \bra{j}$ with $R_0$$(R_1)$ as decrement (increment) shift operators responsible for anti-clockwise (clockwise) movement of the walker on a cyclic graph. The unitary time evolution operator of the quantum walk is given as, \begin{equation}
    U_N = S\cdot (I_N \otimes C_2), \label{e4}
\end{equation} where $I_N$ is an \( N\times N\) identity matrix and $U_N$ is a \( 2N\times 2N\)  circulant matrix given as~\cite{Dukes} ,
    \begin{align}
U_N(r,a,b) &= \text{CIRC}_N \Bigg(
\begin{bmatrix} 0 & 0 \\ 0 & 0 \end{bmatrix}_{0},\  
\begin{bmatrix} \sqrt{r} & \sqrt{1-r} e^{i a} \\ 0 & 0 \end{bmatrix}_{1},\  
\cdots \Bigg.  \notag \\ 
& \quad \Bigg. ,\begin{bmatrix} 0 & 0 \\ \sqrt{1-r} e^{i b} & -\sqrt{r} e^{i(a+b)} \end{bmatrix}_{N-1}
\Bigg).
\label{e5}\end{align} The walker applies $U_N$ repeatedly on its initial state to reach the final state. For $t$ steps, $ |\psi(t) \rangle = U_N^t
 |\psi(t=0)\rangle $. If the walker returns to its initial state after $t=T$ steps, that is, \begin{equation} \ket{\psi(t=T)} = U_N^T |\psi(0)\rangle = |\psi(0) \rangle,
 \label{e6}\end{equation} then the walk is said to be periodic, that is, ordered with period $T$. Let \{$|x_i\rangle$\} be the eigenvectors of $U_N$ with $ \lambda_i $ being the eigenvalues. The initial state $\ket{\psi(0)}$ can be represented as $|\psi(0)\rangle  = \sum_{j=1}^{2N}\alpha_j|x_j\rangle $ where $\alpha_j$ is the amplitude of the $jth$ eigen vector. Applying $U_N$, $t$ times on the initial state, we get, \begin{equation} U_N^t |\psi(0)\rangle = \sum_{j=1}^{2N}\alpha_j\lambda_j^t|x_j\rangle .\label{e7}\end{equation} From Eqs.~(\ref{e6}) and (\ref{e7}) we get,  \begin{equation} U_N^T = I_{2N},  \text{or} \lambda_j^T = 1,\forall\; 1\leq j \leq 2N, \label{e8}\end{equation} which is the condition of periodicity. If $U_N$ satisfies Eq.(\ref{e8}), it generates an ordered or periodic quantum walk with period T. Otherwise, the quantum walk is chaotic~\cite{treganna}. Further, whether the walk is chaotic or not can also be determined via the Lyapunov exponent ($\alpha$)~\cite{panda,lyapunov}. A positive value of the Lyapunov exponent, i.e., $\alpha>$ 0, implies the quantum walk is chaotic, while if $\alpha =$ 0 then it’s periodic or non-chaotic, for example, with quantum walk on a 3-cycle graph, see Ref.~\cite{panda}. The calculation of the Lyapunov exponent for chaotic and periodic walks in our DTQW model with an example 4-cycle graph is described in Appendix~\ref{5}. \\We emphasize that, unlike classical systems, quantum dynamics governed by a unitary operator does not exhibit chaos in the conventional sense through exponential divergence of trajectories, as unitarity preserves the norm and overlap between quantum states. Nevertheless, quantum chaos can still emerge through aperiodic evolution of state amplitudes in Hilbert space. In bounded systems such as cyclic quantum walks, chaotic behavior manifests through irregular probability distributions~\cite{Dukes,treganna}, sensitivity to the choice of coin parameters~\cite{panda}, non-zero Lyapunov exponents~\cite{lyapunov}(see Appendix~\ref{5}), and the growth of entanglement entropy~\cite{entang}, even though the evolution remains unitary. Correspondingly, spectral statistics often display a transition from Poisson to Wigner–Dyson level-spacing distributions, which serves as another characteristic of quantum chaotic behavior~\cite{Chaos_book}. \\In contrast, for DTQWs on infinite lattices, chaos must be characterized through information-theoretic and dynamical indicators rather than spectral confinement. In such unbounded systems, signatures of chaotic dynamics include the absence of temporal recurrences and extreme sensitivity of the evolved state to infinitesimal perturbations in the coin or initial condition, quantified via Lyapunov exponents~\cite{61}, which also characterize bounded systems. Other manifestations include the rapid, near-linear growth of position–coin entanglement entropy~\cite{60} and the growth and eventual saturation of out-of-time-order correlators (OTOCs), capturing the scrambling of quantum information. The irregular, quasi-random spreading of local operators under the walk unitary, analogous to the quantum butterfly effect, provides additional evidence of chaotic behavior~\cite{59}. Thus, while classical chaos originates from exponential divergence of trajectories in phase space, quantum chaos in DTQWs, whether bounded or unbounded, arises from the complex, non-recurrent evolution of probability amplitudes, heightened sensitivity to control parameters, and the emergence of universal spectral and entanglement signatures, all within the strictly unitary framework of quantum dynamics.\\The periodic nature of the walk is determined by the coin operator employed. For instance, the widely used $Hadamard$ coin ($H=C_2(r=1/2,a=0,b=0)$) is ordered (periodic) for even cyclic graphs such as $4$ and $8$-cycle graphs with period $8$ and $24$ respectively, while chaotic in odd cycle graphs such as $3$ and $5$-cycles. Coins like $R_3 = C_2(r=2/3,a=0,b=0),C = C_2(r=\frac{5-\sqrt{5}}{6},a=0,b=0)$ are periodic for $3$-cycle graphs with period $8$ and $10$ respectively. Ref.~\cite{Dukes} enlists the parameters for coins generating ordered or periodic walks for various cyclic graphs, both even and odd . 
 
\subsection{Parrondo's paradox via quantum walks}\label{2b}
As described earlier, Parrondo's paradox refers to situations where a combination of losing strategies gives rise to a winning outcome. Within the framework of quantum walks, this translates to the possibility that a deterministic combination of individually chaotic walks results in a periodic walk, thereby illustrating the emergence of order from chaos. In Ref.~\cite{panda}, a deterministic combination of unitary coin sequences of chaotic walks yielded periodic quantum walks on cyclic graphs. Let $A = U_N(r_1,a_1,b_1)$ and $B = U_N(r_2,a_2,b_2)$ be two unitary operators that result in chaotic quantum walks, where $U_N$ is defined by Eq.(\ref{e4}). Sequences like $ABAB...$, $AABAAB...$, $ABBABB...$, $AABBAABB...$ were examined (see, Appendix~\ref{6}), out of which the $AABB...$ combination of unitary operators yielded a periodic walk. Our motivation in this work is to implement such Parrondo strategies($AABB..$) on NISQ quantum computers by quantum circuit implementations of DTQWs on arbitrary cyclic graphs (even or odd), to visualize the emergence of order from chaos, and estimate how well the circuits perform in a real noisy medium. 
\section{Quantum Circuit implementing DTQW on Cycles}\label{3}
\subsection{Even Cycle} \label{3a}
To map a problem into a quantum computer, one needs a quantum circuit that computes the set of unitary operations via quantum gates laid onto a set of physical qubits. In order to implement an even $(N = 2^n)$ cycle DTQW on a quantum computer, $n+1$ qubits are required, where $n$ qubits encode the position and an extra qubit encodes the coin state of the walker. The states of the walker are encoded in binary form. An arbitrary state of the walker is represented in terms of physical qubits as, \begin{equation}
|s_c\rangle |j_p\rangle = |q_n^c\rangle|q_{n-1}^pq_{n-2}^p...q_0^p\rangle, \label{e9}\end{equation} where $|q_n^c\rangle$ represents the qubit encoding the coin state $|s_c\rangle$ and $|q_{n-1}^pq_{n-2}^p...q_0^p\rangle$ refers to the qubits encoding the position state of the walker $|j_p\rangle$. Some of the methods of circuit implementations of DTQWs on even cycle graphs are reported in Ref.~\cite{DW,SK,Luca}. A quantum circuit is considered efficient, that is, utilizes minimum resources, if it efficiently implements the unitary-time evolution operator defined in Eq.(\ref{e4}). This implies the efficient application of the conditional shift operator as described in Eq.(\ref{e3}). Ref.~\cite{DW} describes the shift of the walker via multi-qubit generalized controlled NOT operations, accounting for the shift in either direction. Eq.(\ref{e5}) describes the circulant unitary-time evolution operator for cyclic DTQW, which, in general, is diagonalized by the commensurate Fourier matrix tool (see, Ref.~\cite{Dukes}). Here, we note that the increment (decrement) operators, $R_1(R_0)$, are circulant matrices too, hence can be diagonalized for simplification of shift implementation on a quantum circuit. Although the controlled increment-decrement scheme was among the earliest circuit-based implementations of DTQWs on cyclic graphs, it is not efficient due to its reliance on numerous multi-qubit gates, which are prone to errors and inhibit parallel gate execution in current NISQ computers. In contrast, Refs.~\cite {SK} and~\cite{Luca} adopt the quantum Fourier transform (QFT) approach to diagonalize the shift operator, significantly reducing the number of multi-qubit gates and allowing for more efficient, parallel execution of gate layers. To investigate the transition from chaotic to ordered dynamics in cyclic graphs, we adopt the QFT-based scheme for implementing the shift operator.

\begin{figure}
    \centering
    \includegraphics[width=0.9\linewidth]{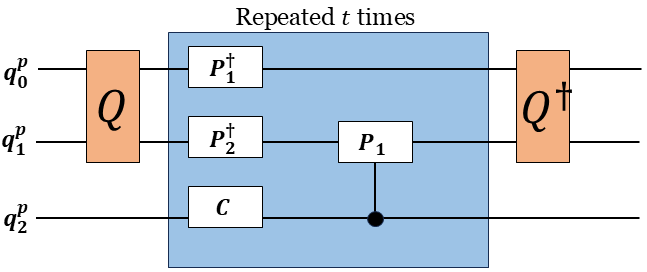}
    \caption{Quantum circuit representation of the $4$-cycle DTQW dynamics. $P$ gates refer to the phase gates (rotation) that are involved in the shift operation, and $C$ is the coin operator.}
    \label{fig:quantum_circuit}
\end{figure}

%In order for a quantum circuit to execute DTQW efficiently, efficient application of the unitary time evolution operator described as in Eq. (\ref{e4}) is required thereby implying the efficient implementation of the conditional shift operator Eq.$(3)$. In Sec.$A$, we described the block diagonalization of the unitary operator $U_n$ by commensurate Fourier matrix tool, we can thus employ the quantum Fourier transform$(QFT)$ of the conditional shift operator $S$, instead. 
From Eq.$(3)$, the shift operator, \begin{equation} S = \begin{pmatrix} R_0& \mathbf{0}\\ \mathbf{0}& R_1
\end{pmatrix},\label{e10}\end{equation} where $\textbf{0'}$s are $N\times N$ null matrices.
For a 4-cycle graph, \begin{equation} R_0 = \begin{pmatrix}
    0& 1& 0& 0 \\
    0& 0& 1& 0 \\
    0& 0& 0& 1 \\
    1& 0& 0& 0 
    \end{pmatrix}  \text{ and}\text{ }R_1 = \begin{pmatrix}
    0& 0& 0& 1 \\
    1& 0& 0& 0 \\
    0& 1& 0& 0 \\
    0& 0& 1& 0 \end{pmatrix}, \label{e11} \end{equation} 
which are diagonalized by the quantum Fourier transform matrix Q (see, Appendix A of Ref.~\cite{Luca}); also, see Table~\ref{t1} for position transitions based on the coin state of the walker. QFT changes the computational basis as, \begin{equation}
Q :  |j\rangle \xrightarrow{} \frac{1}{\sqrt{N}} \sum_{k = 0}^{N-1}\omega_N^{jk}|k\rangle ,\label{e12}\end{equation} where $\omega_N^{jk} = e^{2\pi i/N}$ and in matrix form , $Q = \omega_N^{jk}/\sqrt{N}$. For 4-cycle graph the QFT matrix $Q$ is given by, \begin{equation}
     Q = \frac{1}{2}\begin{pmatrix}
            1& 1& 1& 1\\
            1& \omega& \omega^2&\omega^3\\
            1& \omega^2& \omega^4&\omega^6\\
            1& \omega^3& \omega^6&\omega^9\\
            \end{pmatrix},
    \end{equation} where $\omega = e^{2\pi i/4}$. The inverse quantum Fourier transform is denoted as $Q^\dagger$ with $QQ^\dagger = Q^\dagger Q = I$. Hence, \begin{equation} R_0 = Q^\dagger \mathcal{P^\dagger} Q \text{ }\text{and}\text{ } R_1 = Q^\dagger \mathcal{P} Q, \label{e13}\end{equation} where $\mathcal{P(P^\dagger)}$ are the diagonalized decrement and increment operators respectively.
\begin{table}[ht!]
    \centering
    \renewcommand{\arraystretch}{1.1}
    
    \begin{tabular}{|c|c|c|c|}
        \hline
        \textbf{Model} & \textbf{Position} & \textbf{Coin State $\ket{0}$} & \textbf{Coin State $\ket{1}$}\\
        \hline
        \multirow{4}{*}{4-cycle DTQW } & 0& 3 & 1\\
        \cline{2-4}
        & 1& 0&2\\ \cline{2-4}
        & 2 &1 &3\\ \cline{2-4}
        & 3&2 &0 \\
        \hline
    \end{tabular}
    \caption{Position transitions of the walker during the shift operation based on the coin states for 4-cycle graph.}
    \label{t1}
\end{table}%
\\In case of 4-cycle, \begin{equation} \begin{split} \mathcal{P} = \begin{pmatrix}
    1& 0 & 0& 0 \\
    0& \omega& 0& 0\\
    0& 0& \omega^2& 0\\
    0& 0& 0& \omega^3\end{pmatrix}  = \begin{pmatrix} 1\ 0 \\
    0\ \omega^2 \end{pmatrix} \otimes \begin{pmatrix} 1\ 0\\
    0\ \omega \\ \end{pmatrix} \\
        = P_1(\theta_1) \otimes P_2(\theta_2),
    \end{split}
    \label{e14}
    \end{equation}
    with $\omega = e^{2\pi i /4}$ and $P_i(\theta_i)$ are the phase gates which account for the shift of the walker via rotations given as, \begin{equation}
        P_i(\theta_i) = \begin{pmatrix}
            1& 0\\ 0& e^{i\theta_i}
        \end{pmatrix}.
         \label{e15}
    \end{equation}
       $\mathcal{P}$ acts only on the position qubits such that, $\mathcal{P}|q_1q_0\rangle = P_1|q_0\rangle \otimes P_2|q_1\rangle$ . Thus, the diagonalized conditional shift operator is given as, \begin{equation} \mathcal{S} = (I_c\otimes Q)S(I_c\otimes Q^\dagger) = |0\rangle\langle0| \otimes \mathcal{P^\dagger} + |1\rangle\langle1|\otimes \mathcal{P}, \label{e16}\end{equation} where $\mathcal{S}$ is the diagonalized shift operator. Since QFT acts only on the position space, one can write $(C\otimes I_p)$ as $(I_c \otimes Q^\dagger)(C\otimes I_p)(I_c\otimes Q)$, hence, the unitary time evolution operator has the form~\cite{Luca}, \begin{equation} U^t = (I_c\otimes Q^\dagger)[\mathcal{S}(C\otimes I_p]^t(I_c\otimes Q). \label{e17}\end{equation} Using the property $QQ^\dagger = Q^\dagger Q = I $, leads to the usage of only one quantum Fourier transform and one inverse quantum Fourier transform in the circuit, hence an optimized, efficient circuit, see Fig.~\ref{fig:quantum_circuit}. Thus, \begin{equation}  U^t = (I_c\otimes Q^\dagger)[(|0\rangle\langle0| \otimes I_p + |1\rangle\langle1|\otimes \mathcal{P}^{2})(C\otimes \mathcal{P^\dagger})]^t(I_c\otimes Q). \label{e18}\end{equation}
    As current NISQ quantum computers are restricted to limited connectivity, with the choice of high-fidelity adjacent physical qubits in real quantum hardware, one can obtain an improved performance of a quantum circuit with fewer resources.

\subsection{NISQ implementation of Parrondo's strategy on 4-cycle graph}\label{3b}
\begin{figure}[h]
    \centering
    \includegraphics[width=1\linewidth]{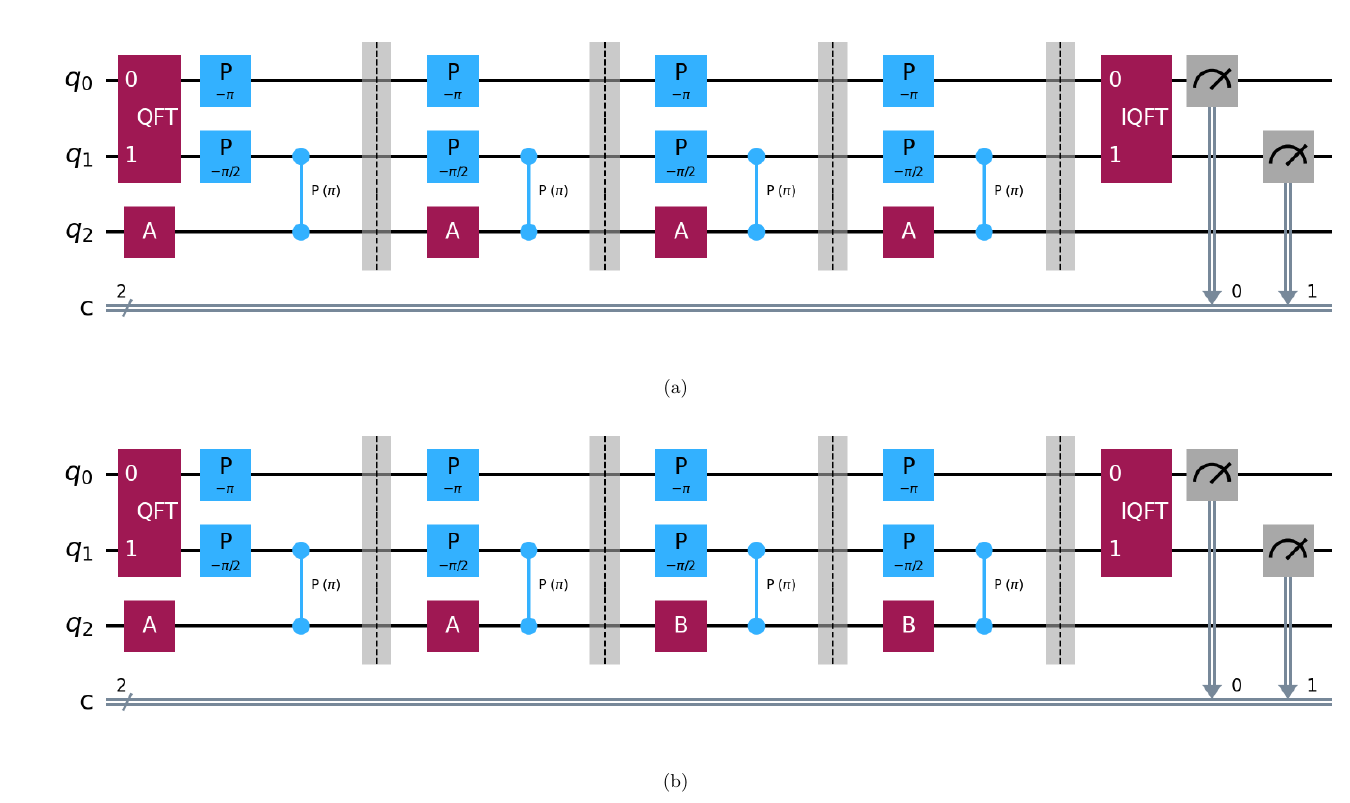}

    \caption{Quantum circuits implementing the sequences (a)$AAAA..$(or $BBBB..$) and (b)$AABB..$ on $4$-cycle for $4$ time steps designed within Qiskit framework.}
    \label{fig:quantum-circuits}
\end{figure}

\begin{figure}
    \centering
    \includegraphics[width=0.7\linewidth]{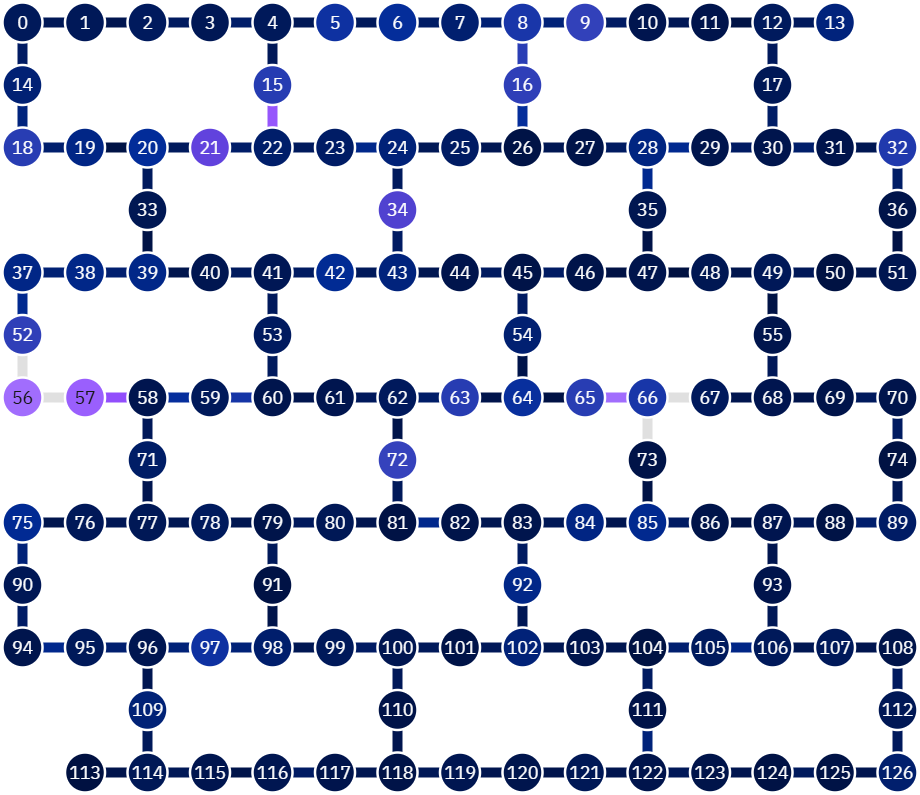}
    \caption{Qubit connectivity of \texttt{ibm\_sherbrooke}(Image Source: IBM Quantum~\cite{ibm} Accessed on December 2024).}
    \label{fig:connectivity}
\end{figure}

In order to explore Parrondo's paradox in real quantum hardware, we implement the quantum circuits shown in Figs.~\ref{fig:quantum-circuits} for 4-cycle DTQW, in three quantum computers, namely, \texttt{ibm\_sherbrooke v 1.6.10}, \texttt{ibm\_brisbane v 1.1.81}, and \texttt{ibm\_kyiv v 1.20.22}, which are 127-qubit quantum devices under Eagle $r3$ processor. The IBM quantum systems are accessed via IBM cloud services~\cite{ibm}, and the numerical experiments are carried out within the Qiskit framework~\cite{qiskit}. The algorithms can be accessed in the Appendix~\ref{Algorithm}. The qubit connectivity of \texttt{ibm\_sherbrooke} is shown in Fig.~\ref{fig:connectivity} while the connectivity of the other two NISQ devices can be found in ~\cite{ibm}. In the following sections, we describe the transition from chaos to order on 4-cycle graphs using the Parrondo strategy, compute the probability distribution with fidelity, and then visualize the results on a quantum computer.

Throughout this section, the walker is initialized at $|s_c\rangle|j_p\rangle = |0_c0_p\rangle$, that in quantum computer reads, $|q_2q_1q_0\rangle = |000\rangle$ (see, Eq.~(\ref{e9})).

\textbf{Probability Distribution: } In this work, we are interested in the probability distribution of the walker at the initial position, as it determines whether the quantum walk is ordered or not. For this purpose, we perform ideal simulations using \texttt{AerSimulator()}, a local simulator without noise, which is followed by the noisy simulations generated by \texttt{NoiseModel()} in \texttt{qiskit\_aer} to get some information on real dynamics before applying it on a NISQ quantum computer. Then the dynamics of the DTQW on cyclic graphs are realized in NISQ devices. The quantum circuits designed via Qiskit are shown in Figs.~\ref{fig:quantum-circuits} and \ref{fig: AABB ckts}, are first simulated in \texttt{AerSimulator()} and then realized in quantum hardware.
%For this purpose, we perform ideal simulations using \texttt{AerSimulator()}, a local simulator without noise this is followed by the noisy simulations generated by \texttt{NoiseModel()} in \texttt{qiskit\_aer}. Then the dynamics of the DTQW on cyclic graphs are realized in NISQ devices:  \texttt{ibm\_sherbrooke}, \texttt{ibm\_brisbane}, and \texttt{ibm\_kyiv}. 

To compare probability distributions obtained via ideal simulations and real implementations of DTQW, we calculate the \textit{Hellinger Fidelity}. Let \( P = \{p_k\} \) and \( Q = \{q_k\} \), be two probability distributions, for instance, $P$ can be the probability distribution obtained via ideal simulations in \texttt{AerSimulator()} while $Q$ is the probability distribution realized in real hardware. We calculate the \textit{Hellinger fidelity}~\cite{HF, HF_1} defined as,\begin{equation}
    H(P,Q) = \left[ 1 - h^2(P,Q) \right]^2,
\label{e22}\end{equation}
where $h$ is the Hellinger distance given as:

\begin{equation}
    h(P, Q) = \frac{1}{\sqrt{2}} 
    \sqrt{\sum_{k} \left(\sqrt{p_k} - \sqrt{q_k}\right)^2}.\label{e23}
\end{equation}

The \textit{Hellinger fidelity} is bounded by \( [0,1] \), where 1 indicates that the distributions are equivalent, while 0 implies that they are completely distinct. Higher fidelity (\( >0.5 \)) suggests that the distributions are similar, while a fidelity $>0.95$  suggests that they are almost alike. Since our focus is on the probability dynamics of the position states, we use Hellinger fidelity as it naturally quantifies the closeness between probability distributions obtained via ideal simulations in \texttt{qiskit\_aer}, and actual implementations in NISQ hardware. One can also calculate the Bhattacharya coefficient($BC$)~\cite{BC} to compare the probability distributions $P$ and $Q$, given by, \begin{equation}
    BC = {\sum_{k}\sqrt{p_kq_k}}, \end{equation} we calculate, the Bhattacharya fidelity as, \begin{equation}
        F = BC^2, \label{BF}
    \end{equation} which yields exactly similar results as obtained from the Hellinger fidelity (see, Appendix~\ref{7}).
\\ To implement Parrondo's strategy$(AABB...)$ on a $4$-cycle ($2^2)$, we require $3$ qubits. As discussed in Sec.~\ref{2b}, the coin operators of $A$ and $B$ result in chaotic walks, defined as $A = C_2(r_1 = 0.998489, a_1 = 0, b_1 =0)$ and $B = C_2( r_2 = 0.119545, a_2 = 0 , b_2 = 0)$ respectively (see, Eq. \ref{e2}). The quantum circuits shown in Fig.~\ref{fig:quantum-circuits} are implemented in three quantum devices, \texttt{ibm\_sherbrooke, ibm\_brisbane} and \texttt{ibm\_kyiv} transpiled (converted to the native gate structure of the quantum devices) at the highest level of optimization (level = 3) for $10^5$ shots (times). 

\begin{figure}[h]
    \centering
    \includegraphics[width=1\linewidth]{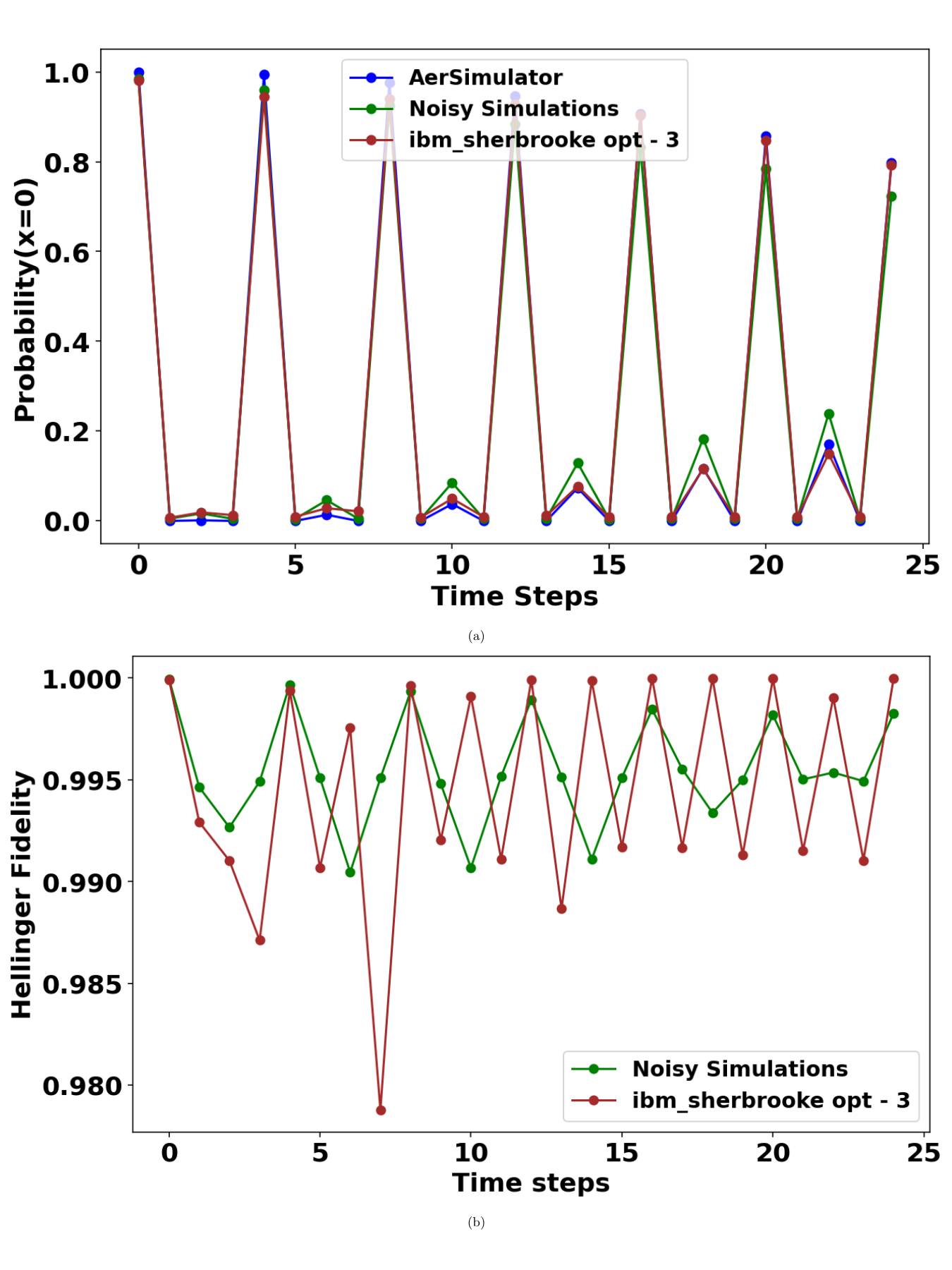}
     
     \caption{Chaotic QW on $4$-cycle graphs for sequence of chaotic coins (a) $AAA$...and the (b) Hellinger fidelity. The circuits are implemented on \texttt{ibm\_sherbrooke} at optimization level 3 for $10^5$ shots with position state being encoded in qubits 46,47, and coin state in qubit 48.}
    
    \label{fig:probs}
\end{figure}%
\begin{figure}[h]
    \centering
    \includegraphics[width=1\linewidth]{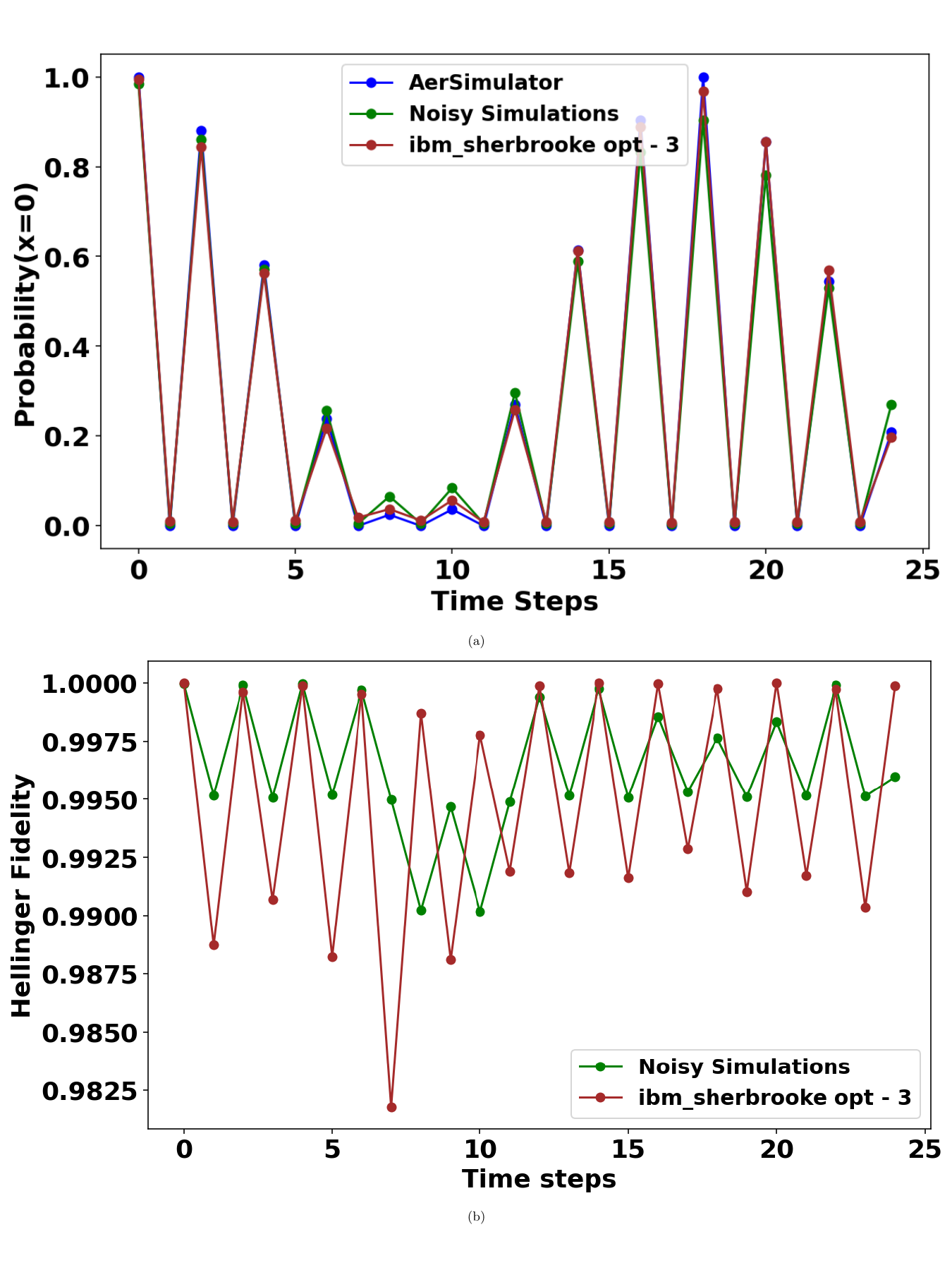}
     \caption{Chaotic QW on $4$-cycle graphs with chaotic coin sequences (a) $BBB$... and (b) Hellinger fidelity. The circuits are implemented on \texttt{ibm\_sherbrooke} at optimization level $3$ for $10^5$ shots with position state being encoded in qubits 46,47, and coin state in qubit 48.}
    
    \label{fig:probs2}
\end{figure}%
\begin{figure}[h]
\centering
    \includegraphics[width=1\linewidth]{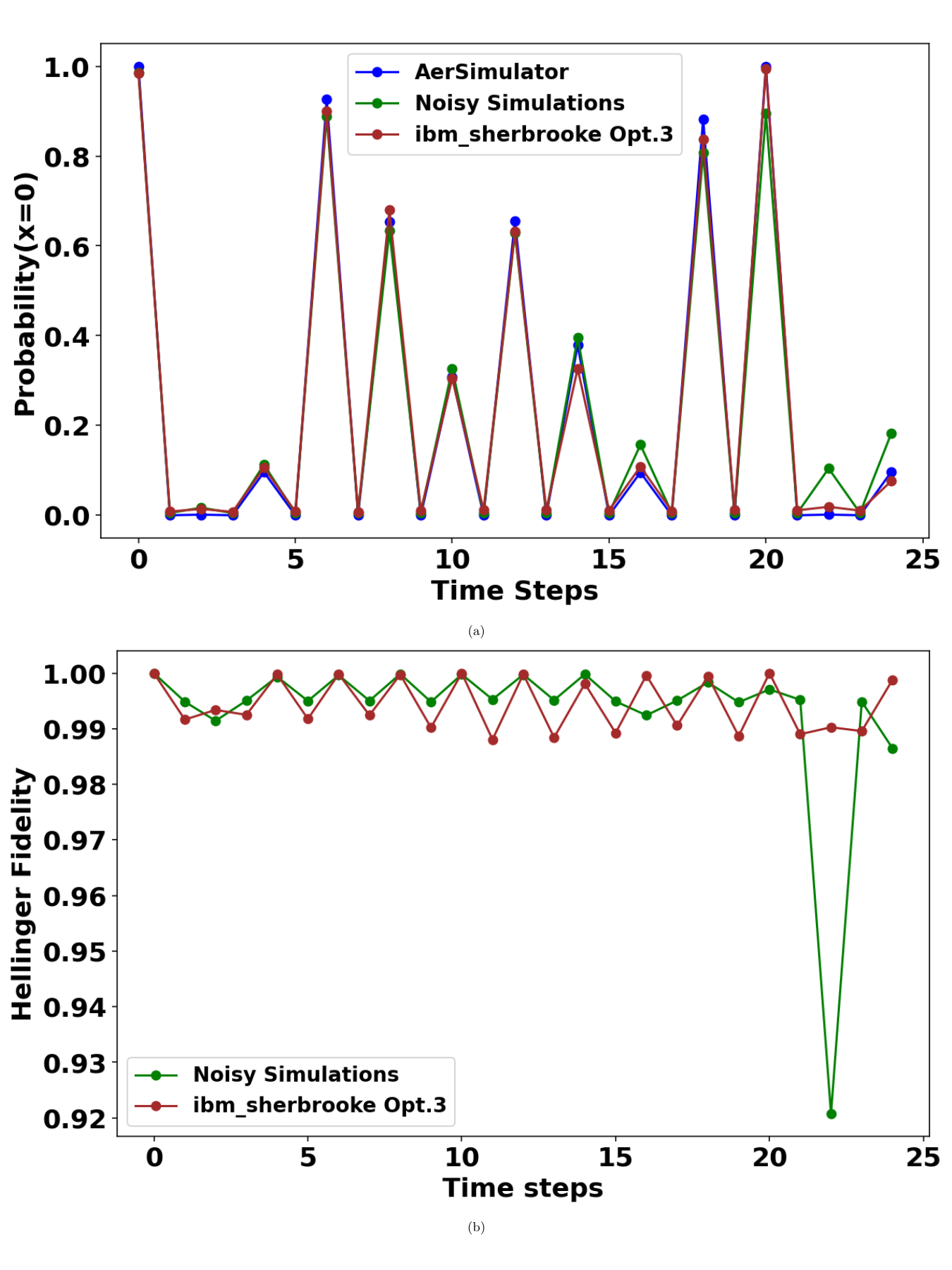}
     \caption{$(a)$ Ordered QW on 4-cycle by Parrondo sequence $AABB...$ with periodicity 20 implemented in \texttt{ibm\_sherbrooke} for 25 time steps with $(b)$ Hellinger fidelity. Results are obtained for $10^5$ shots with position state being encoded in qubits 46,47, and coin state in qubit 48.}
    
    \label{fig: AABB prob}
\end{figure}%
Figs.~\ref{fig:probs} and ~\ref{fig:probs2} show the probability distribution at the initial position and Hellinger fidelity versus time steps of DTQW generated via \texttt{AerSimulator} with and without noise and the real quantum hardware, \texttt{ibm\_sherbrooke} depicting the chaotic nature of the unitary operators with coin sequences $AAA..$ and $BBB...$ along with their Hellinger fidelities. The quantum device, \texttt{ibm\_sherbrooke}, realizes the chaotic nature of unitary coins with $>98\%$ fidelity. Fig.~\ref{fig: AABB prob} shows that the deterministic combination $AABB..$ of the chaotic coins $A$ and $B$ generates an ordered walk with a periodicity of 20, and this agrees well with the theoretical results reported in Ref.~\cite{panda}. The NISQ devices capture the periodicity of the unitary sequence $AABB..$ with fidelity $>95\%$ up to $25$ steps. The actual implementations outperform the noisy simulations generated by \texttt{qiskit\_aer}, mitigating the noise and having better fidelity values. We also find that \texttt{ibm\_sherbrooke} and \texttt{ibm\_kyiv} provide better results than those of \texttt{ibm\_brisbane} (see, Appendix~\ref{results1}) by maintaining fidelity $>98\%$ throughout the walk. Here, we also find that when the circuits are transpiled at the level of optimization (level 3), the depth of the circuit (number of sequential operations of gate layers) becomes shallow, hence, being constant throughout the execution (see, Fig.~\ref{fig: dd plot}). This indicates that this scheme and the level of optimization prove to be efficient for performing DTQW, and thus, visualizing the transition from chaos to order on a quantum computer. We have considered adjacent qubits in the quantum devices in order to eliminate the additional noise introduced during swapping the non-adjacent physical qubits via two-qubit $SWAP$ gates, to perform gate operations.
\subsection{Odd cycle}
\label{3c}
\begin{figure}
    \centering
    \includegraphics[width=1\linewidth]{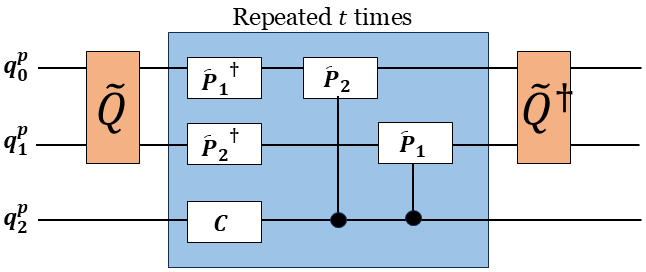}
    \caption{Quantum circuit representation of the $3$-cycle DTQW dynamics. $\tilde{P}$ gates refer to the phase gates (rotation) that are involved in the shift operation (defined in Eq. \ref {e15}), and $C$ is the coin operator.}
    \label{fig:quantum_circuit3}
\end{figure}

Visualizing the transition of chaos to order in an odd cycle $(<2^n)$ DTQW in a quantum computer requires the same number of qubits as the even cycles,i.e., $n$ qubits to encode the position and an additional qubit for the coin state of the walker. As pointed out in ~\cite{DW}, one needs to make trivial alterations to implement a cycle of any size in a quantum circuit. In this work, we present the circuit implementation of DTQW on a $3$-cycle graph; a similar approach can be used for any odd cycle graph.\\
A $3$-cycle graph is a subgraph of the $4$-cycle graph; hence, $2$ qubits are required to encode the position state of the walker. However, $2$ qubits would correspond to $4$ nodes,that is, ${0,1,2,3}$ as position sites. Therefore, to build a $3$-cycle walk, we must keep the last node isolated and unused. The walker moves on the graph by the action of the shift operator (Eq. (\ref{e3})) described in Table~\ref{t11}. 
\begin{table}
    \centering
    \renewcommand{\arraystretch}{1.1}
    
    \begin{tabular}{|c|c|c|c|}
        \hline
        \textbf{Model} & \textbf{Position} & \textbf{Coin State $\ket{0}$} & \textbf{Coin State $\ket{1}$}\\
        \hline
        \multirow{4}{*}{3-cycle DTQW } & 0& {2} & 1\\
        \cline{2-4}
        & 1& 0&2\\ \cline{2-4}
        & 2 &1 &0\\ \cline{2-4}
        & 3&3 &3 \\
        \hline
    \end{tabular}
    \caption{Position transitions of the walker during the shift operation based on the coin states for the 3-cycle graph.}
    \label{t11}
\end{table}%
The decrement and increment operators will now have the form,\begin{equation} R_0 = \begin{pmatrix}
    0& 1& 0& 0 \\
    0& 0& 1& 0 \\
    1& 0& 0& 0 \\
    0& 0& 0& 1 
    \end{pmatrix}  \text{and}\text{ } R_1 = \begin{pmatrix}
    0& 0& 1& 0 \\
    1& 0& 0& 0 \\
    0& 1& 0& 0 \\
    0& 0& 0& 1 \end{pmatrix}, \label{e19}\end{equation} which would be diagonalized by the quantum Fourier matrix (modified for 3-cycle), $\tilde{Q}$ given as, \begin{equation}
\tilde{Q} = \frac{1}{\sqrt{3}} 
\begin{pmatrix}
1 & 1 & 1 & 0 \\
1 & \omega & \omega^2 & 0 \\
1 & \omega^2 & \omega^4 & 0 \\
0 & 0 & 0 & \sqrt{3} \\
\end{pmatrix},
\label{e20}\end{equation} with $\omega = e^{2\pi i/3}$. With these shift operator adjustments and the optimized QFT approach (from Eq. (\ref{e13})- (\ref{e18})), we implement the dynamics of DTQW on a $3$-cycle graph and propose a quantum circuit given in Fig.~\ref{fig:quantum_circuit3}.

\subsection{NISQ implementation of Parrondo's strategy in 3-cycle}\label{3d}
\begin{figure}[H]
    \centering
    \includegraphics[width=0.75\linewidth]{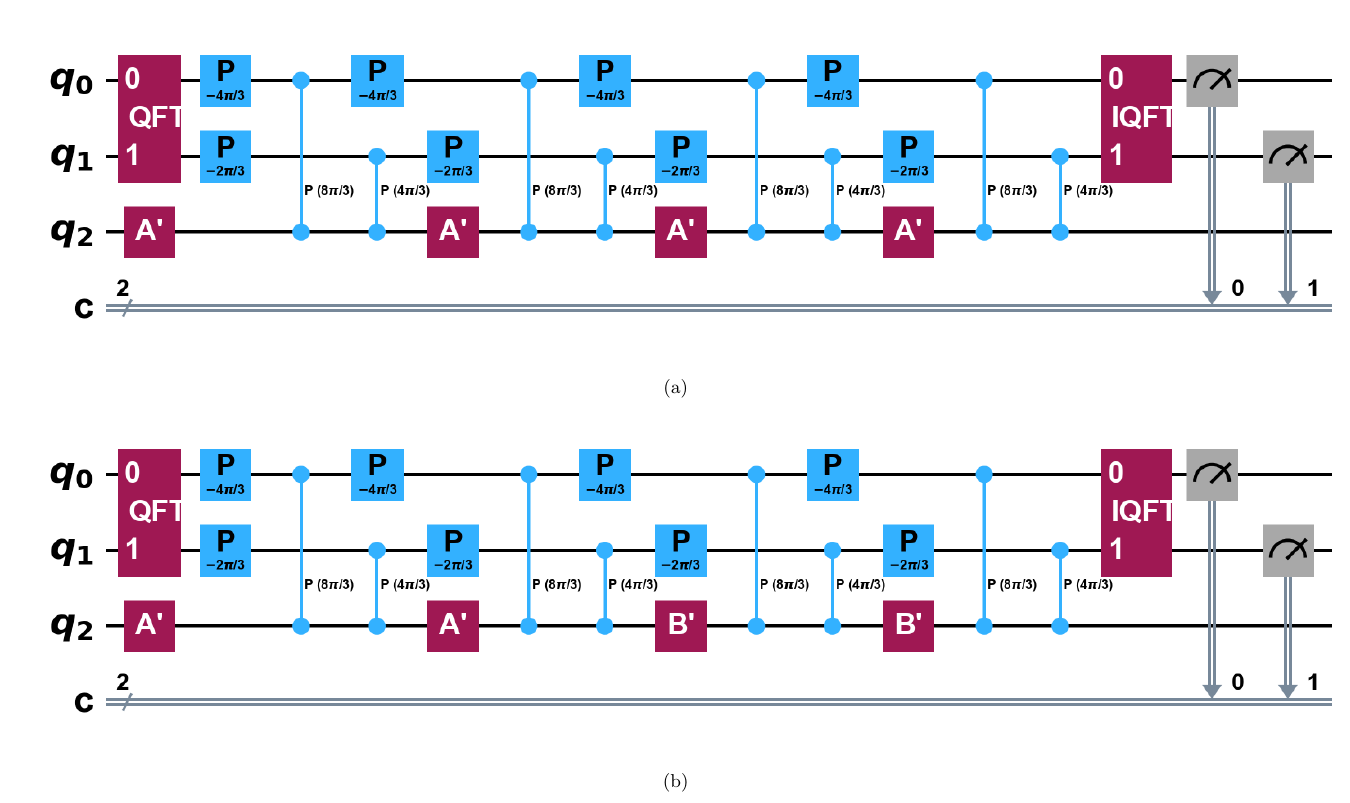}

    \caption{Quantum circuits implementing the Parrondo's strategies (a)$A'A'A'A'..$(or $B'B'B'B'..$) and (b)$A'A'B'B'..$ on $3$-cycle for $4$ time steps designed within Qiskit framework.}
    
    \label{fig: AABB ckts}
\end{figure}

In order to visualize the transition from chaos to order on 3-cycle graphs, we follow a similar approach as for the 4-cycle graphs. First, we implement the circuits shown in Fig.~\ref{fig: AABB ckts} within the Qiskit framework with and without noise, followed by realizing them on real NISQ hardware at the highest level of optimization (level 3). We compute the probability distributions corresponding to the chaotic and ordered QWs obtained via ideal simulations and actual implementations and then compare them by calculating their Hellinger fidelities. As the coins used to implement 3-cycle dynamics would be different from those of 4-cycle, we now denote the chaotic coins as $A'$ and $B'$ respectively. 

\begin{figure}[H]
    \centering
    \includegraphics[width=0.9\linewidth]{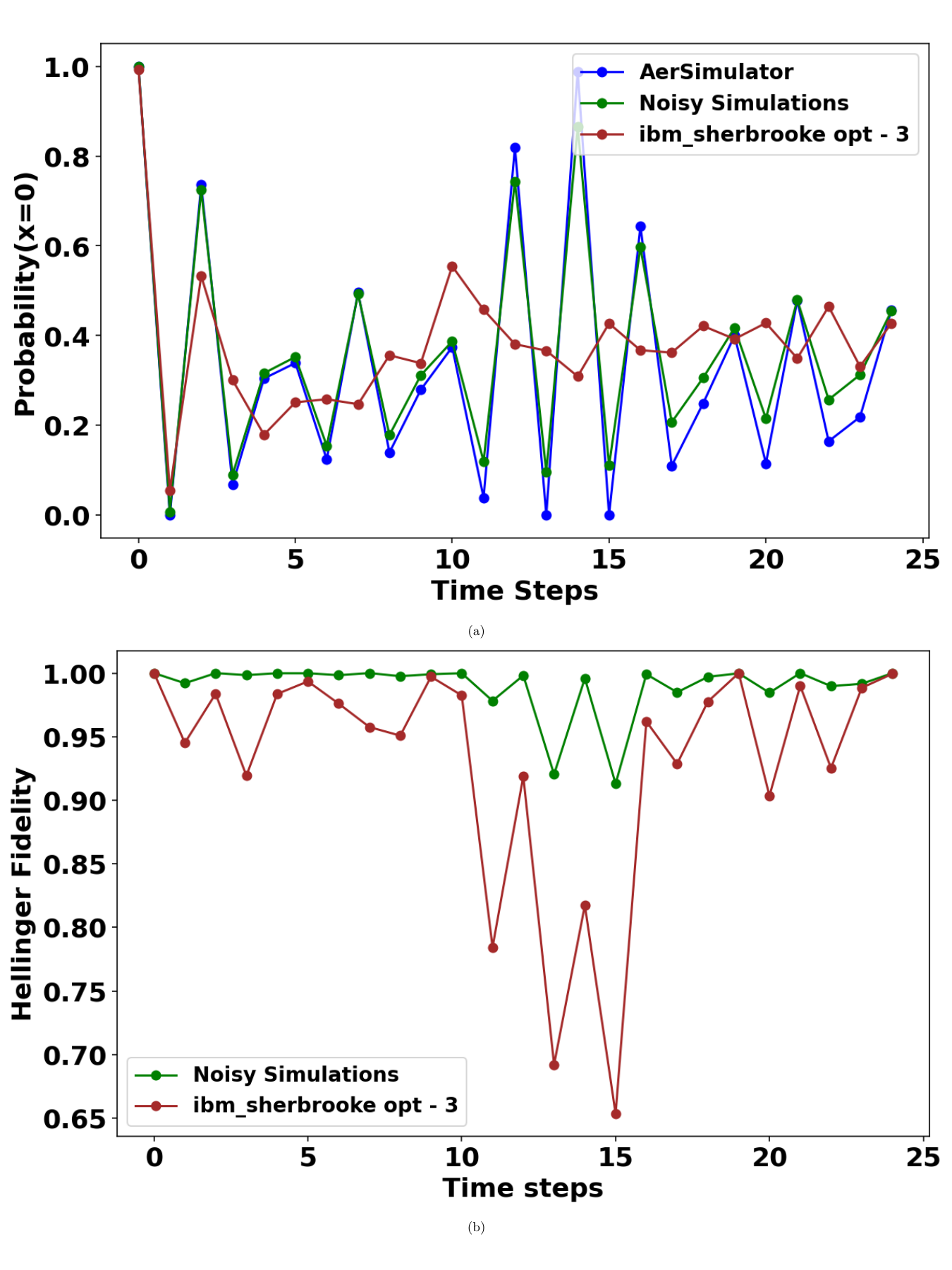}

    \caption{Chaotic QW on $3$-cycle graphs with (a) $A'A'A'$...with  (b) Hellinger fidelity. The circuits are implemented on \texttt{ibm\_sherbrooke} at optimization level 3 for $10^5$ shots with position state being encoded in qubits 69,70, and coin state in qubit 74.}
    
    \label{fig: A}
\end{figure}%

\begin{figure}
    \centering
    \includegraphics[width=1\linewidth]{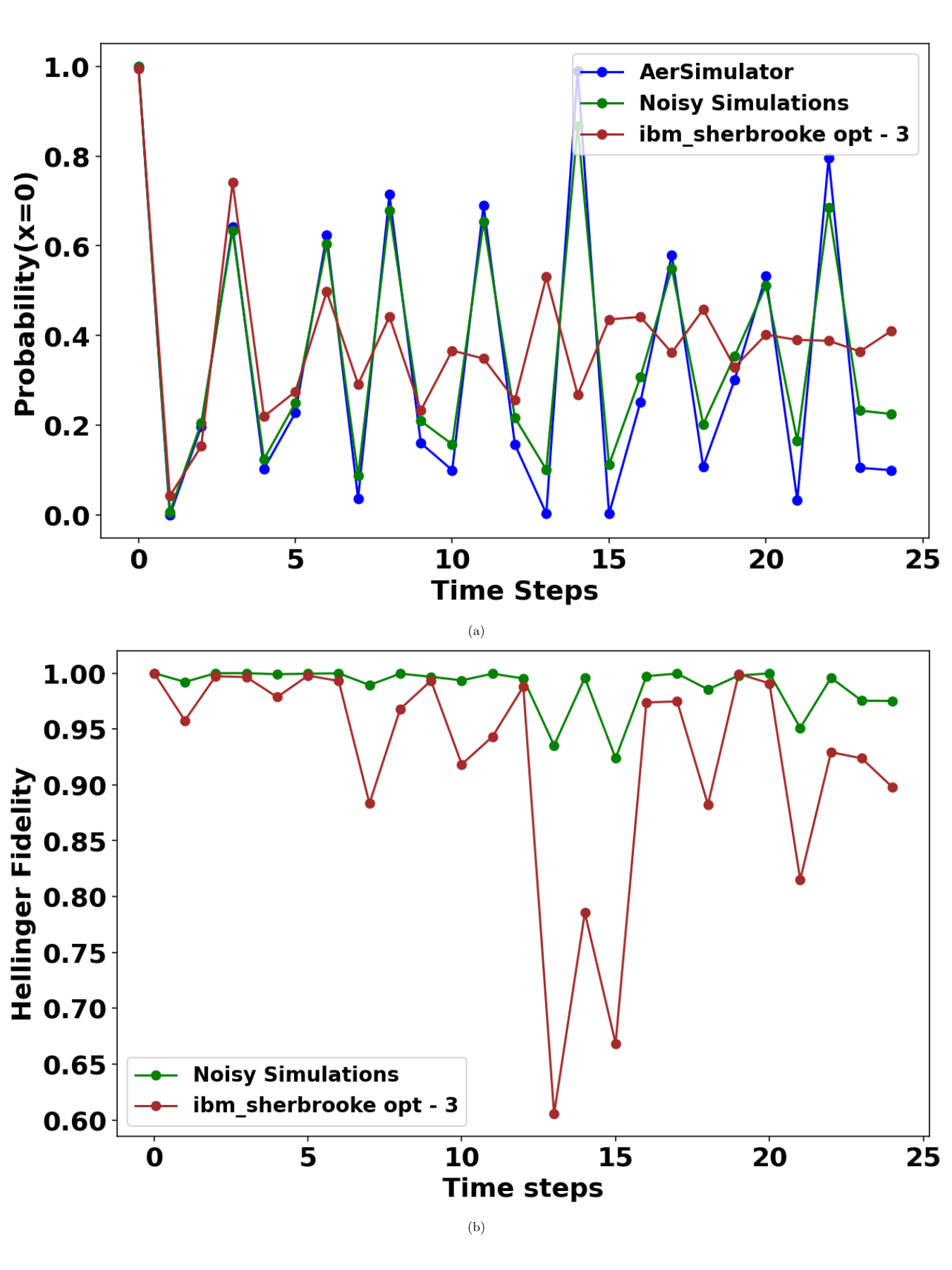}
    \caption{Chaotic QW on $3$-cycle graphs with (a) $B'B'B'$...with  (b) Hellinger fidelity. The circuits are implemented on \texttt{ibm\_sherbrooke} at optimization level 3 for $10^5$ shots with position state being encoded in qubits 69,70, and coin state in qubit 74.}
    
    \label{fig: B}
\end{figure}%

\begin{figure}[h!]
    \centering
    \includegraphics[width=1\linewidth]{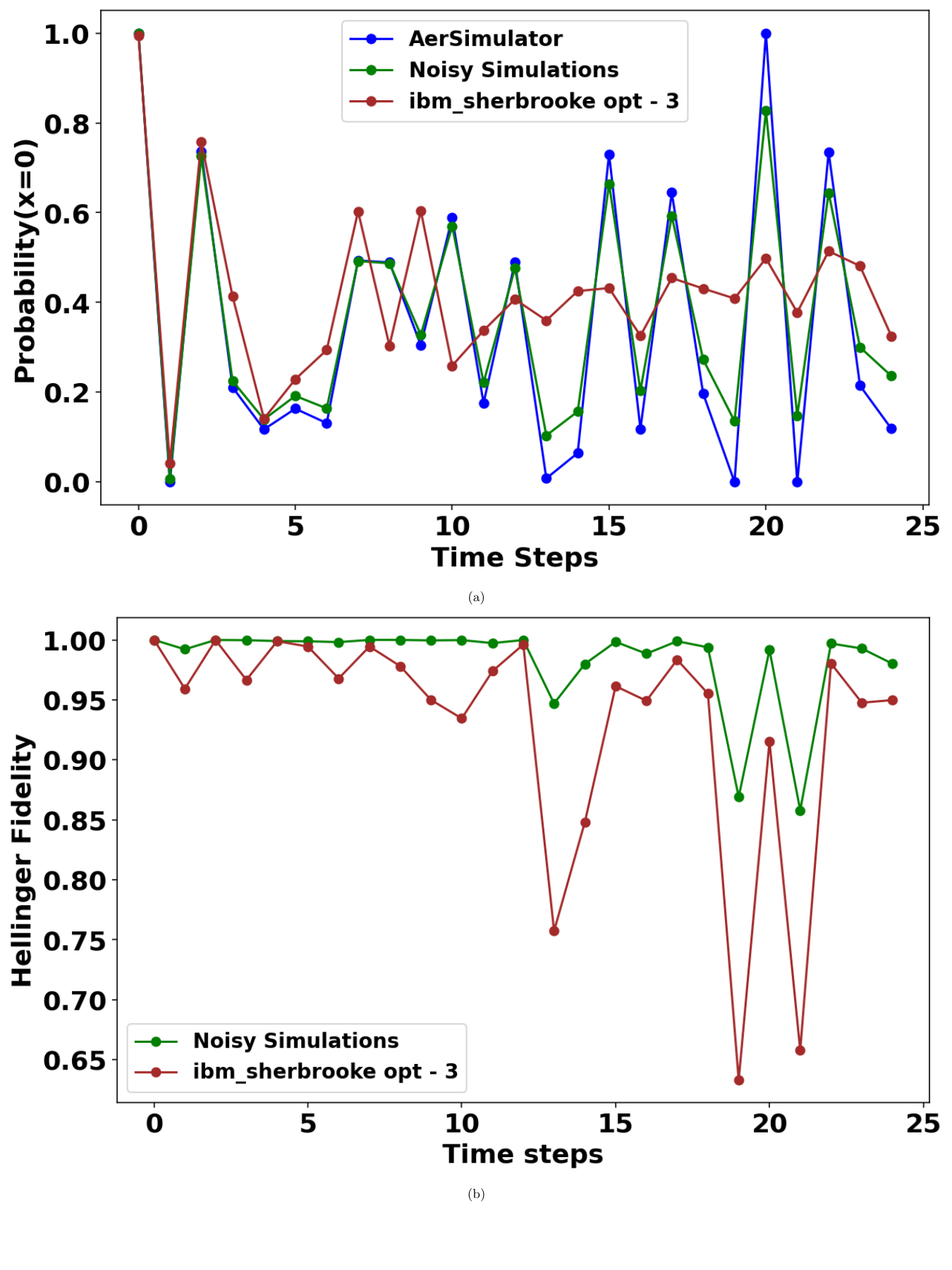}
    
    \caption{$(a)$ Ordered QW on 3-cycle by Parrondo sequence $A'A'B'B'...$ with periodicity 20 implemented in \texttt{ibm\_sherbrooke} for 25 time steps with $(b)$ Hellinger fidelity. Results are obtained for $10^5$ shots with position state being encoded in qubits 69,70, and coin state in qubit 74.}
    
    \label{fig: AABB}
\end{figure}

In order to visualize the Parrondo's strategy $A'A'B'B'...$ for $3$-cycle, we set the parameters of the chaotic coins $A'$ and $B'$ as $C_2(r'_1 = 0.264734,a'_1 = 0,b'_1=0)$ and $C_2(r'_2 = 0.801571,a'_2 = 0,b'_2=0)$ respectively. The walker is initialized at $|q_2q_1q_0\rangle = |000\rangle$. The chaotic nature of walks due to coins $A'$ and $B'$ and the ordered walk produced due to the deterministic combination $A'A'B'B'..$ are shown in Fig.~\ref{fig: A},~\ref{fig: B},~\ref{fig: AABB}. However, using the same set of qubits (46, 47, and 48) as that of the 4-cycle does not provide higher fidelity. Hence, we encode our quantum circuit with a different set of qubits (69, 70, and 74). We further discuss the fidelity values with the choice of qubits in the Appendix~\ref{results}. The periodicity is realized by the AerSimulator(); however, it is not observed in the NISQ implementation. Even the chaotic nature of the operators $A'$ and $B'$ is not well realized by the NISQ device after a certain time steps. In all three cases (both chaotic and ordered), a saturation in the probability distribution is observed with Hellinger fidelities being $>65\%$ for $A'A'A'..$,$>60\%$ for $B'B'B'..$ and $\sim65\%$ for $A'A'B'B'$ respectively.
This discrepancy might be due to the modification of QFT (IQFT) in the $3$-cycle graphs defined in Eq.(\ref{e20}). In Ref.~\cite{Dinesh}, it is explained that due to disorder in position, the probability distribution saturates to a fixed value after a few steps. Similarly, in this case, QFT accounts for the shift operation, which implies that modification in the position via QFT leads to a position disorder in the quantum system, resulting in a saturation of probability. We also observe that in comparison to the $4$-cycle graphs, the number of controlled two-qubit gates has also increased here, which also accounts for the observed fidelity issues in the $3$-cycle graphs.
\subsection{Depth and dynamical decoupling in quantum circuits}\label{DD}
\begin{figure}
    \centering
    \includegraphics[width=1\linewidth]{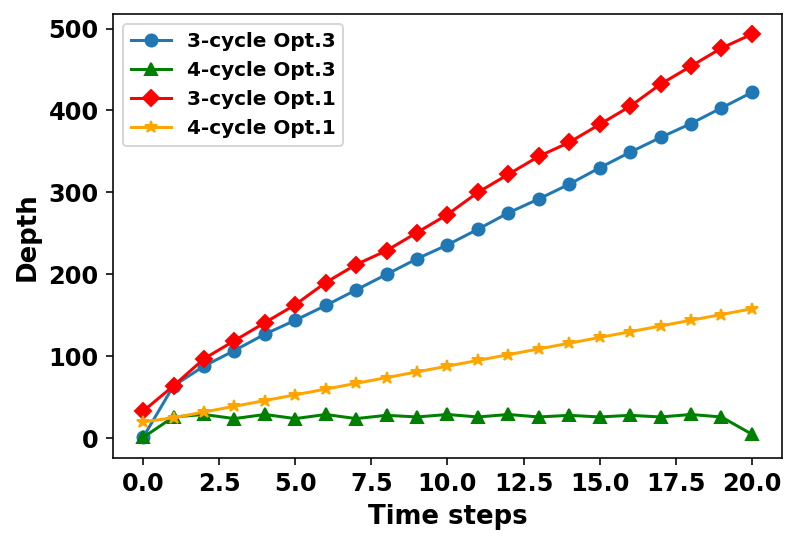}
    \caption{Depth of quantum circuits for 3 and 4-cycle graphs implemented in \texttt{ibm\_sherbrooke} at different optimization levels.}
    \label{fig: dd plot}
\end{figure}

The depth of a quantum circuit refers to the number of sequential layers of quantum gates required to complete a specific computation (see, Appendix~\ref{depth} for calculation of depth of a quantum circuit). A circuit is said to be efficient when the layers of gates are executed in parallel, meaning, lesser depth and a shallow circuit. It is evident from Figs.~\ref{fig:quantum_circuit} and ~\ref{fig:quantum_circuit3} that the depth of the circuits increases linearly for both even and odd cycles. When realized on a quantum computer at the lowest level of optimization (level 1), the linear depth is observed in both 3 and 4-cycles; however, when realized at the highest level (level 3) of optimization, the depth of the circuit for the $4$-cycle decreases to a shallow (constant) value while remaining linear for the
$3$-cycle (see, Fig. ~\ref{fig: dd plot}). This linearity in the 3-cycle circuit gathers noise in the system due to the $2$-qubit and single-qubit gates and hence has poor results in the real hardware. The errors also arise when a qubit remains idle, called idle qubit errors, decoherence, and crosstalk with neighboring qubits. To mitigate these, we integrate the 3-cycle quantum circuit with dynamical decoupling pulses (DD). DD sequences, in general, reduce the idle time experienced by qubits~\cite{DD}, thereby decreasing the decoherence noise introduced due to idle qubits. In the current NISQ era, noise remains a barrier for quantum computers to harness their full potential~\cite{Preskill}. Quantum error correction offers a long-term solution; short-term progress is dependent on error mitigation and suppression techniques. Error suppression methods, such as DD, directly reduce noise in the quantum circuits by decoupling the system from its environment by applying a sequence of pulses such that the effective implementation is identity without changing the overall unitary evolution. However, in real quantum hardware, DD can add additional errors and noise. Some well-known DD sequences are $XY4, XX, XpXm$, etc., for example, $XY4$ is a universal DD sequence defined as~\cite{XY4}, \begin{equation}
    XY4 = Y- \delta - X -\delta - Y - \delta - X - \delta, \label{e21}
\end{equation} where $\delta$ is the duration of the free
evolution. $XY4$ sequence can suppress interactions of the system with the environment by employing $\pi$ rotations around the X- and Y-axes described by the Pauli operators X and Y, respectively.

\begin{figure}
    \centering
    \includegraphics[width=1\linewidth]{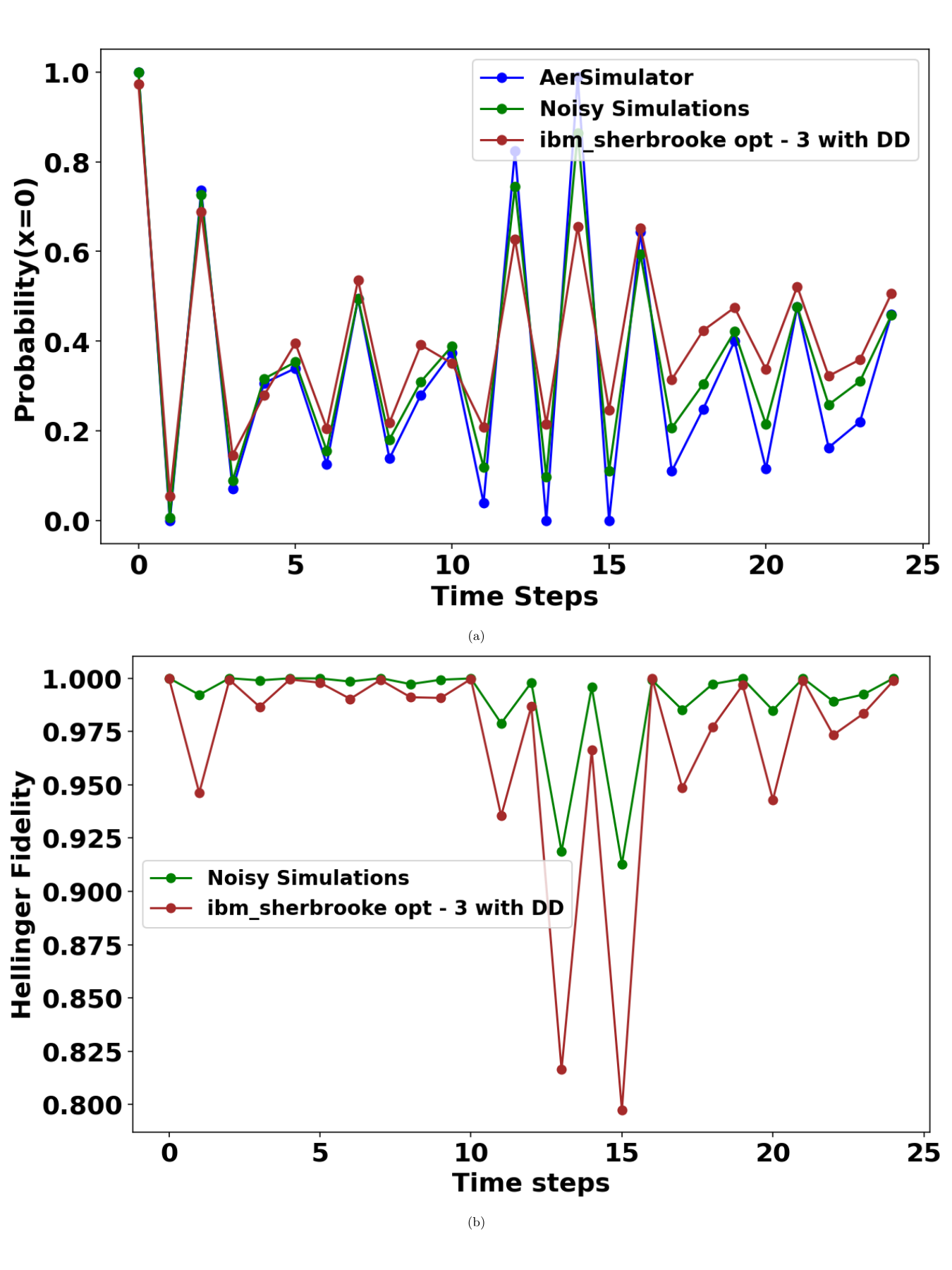}

    \caption{Chaotic QW on $3$-cycle graphs with (a) $A'A'A'$...with  (b) Hellinger fidelity. The circuits are implemented on \texttt{ibm\_sherbrooke} with $XY4$ DD sequence at optimization level 3 for $10^5$ shots with position state being encoded in qubits 69,70, and coin state in qubit 74.}
    \label{fig: A_dd}
\end{figure}

\begin{figure}
    \centering
    \includegraphics[width=1\linewidth]{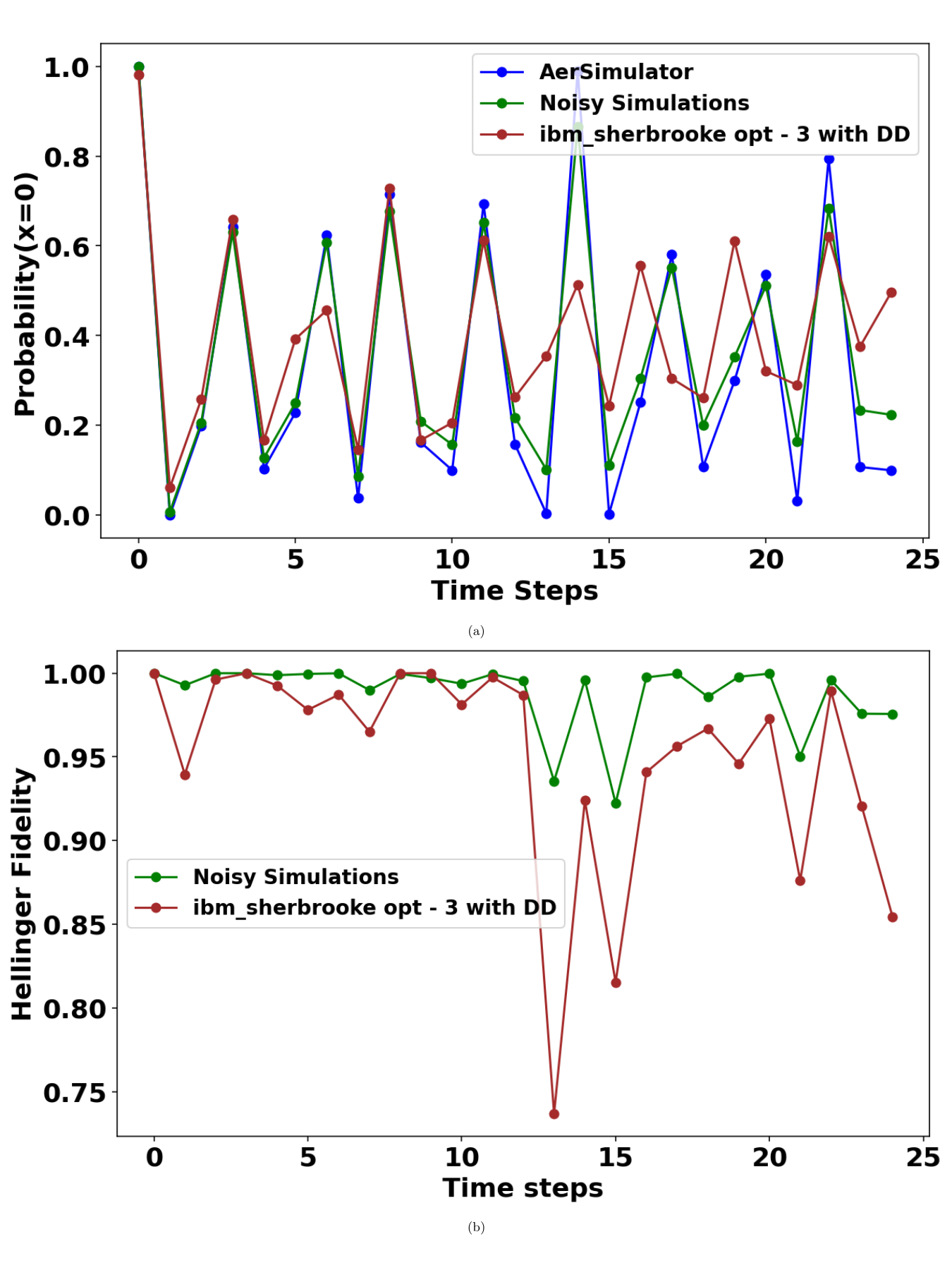}
     \caption{Chaotic QW on $3$-cycle graphs with (a) $B'B'B'$...with  (b) Hellinger fidelity. The circuits are implemented on \texttt{ibm\_sherbrooke} with $XY4$ DD sequence at optimization level 3 for $10^5$ shots with position state being encoded in qubits 69,70, and coin state in qubit 74.}
    
    \label{fig: B_dd}
\end{figure}

\begin{figure}
    \centering
    \includegraphics[width=1\linewidth]{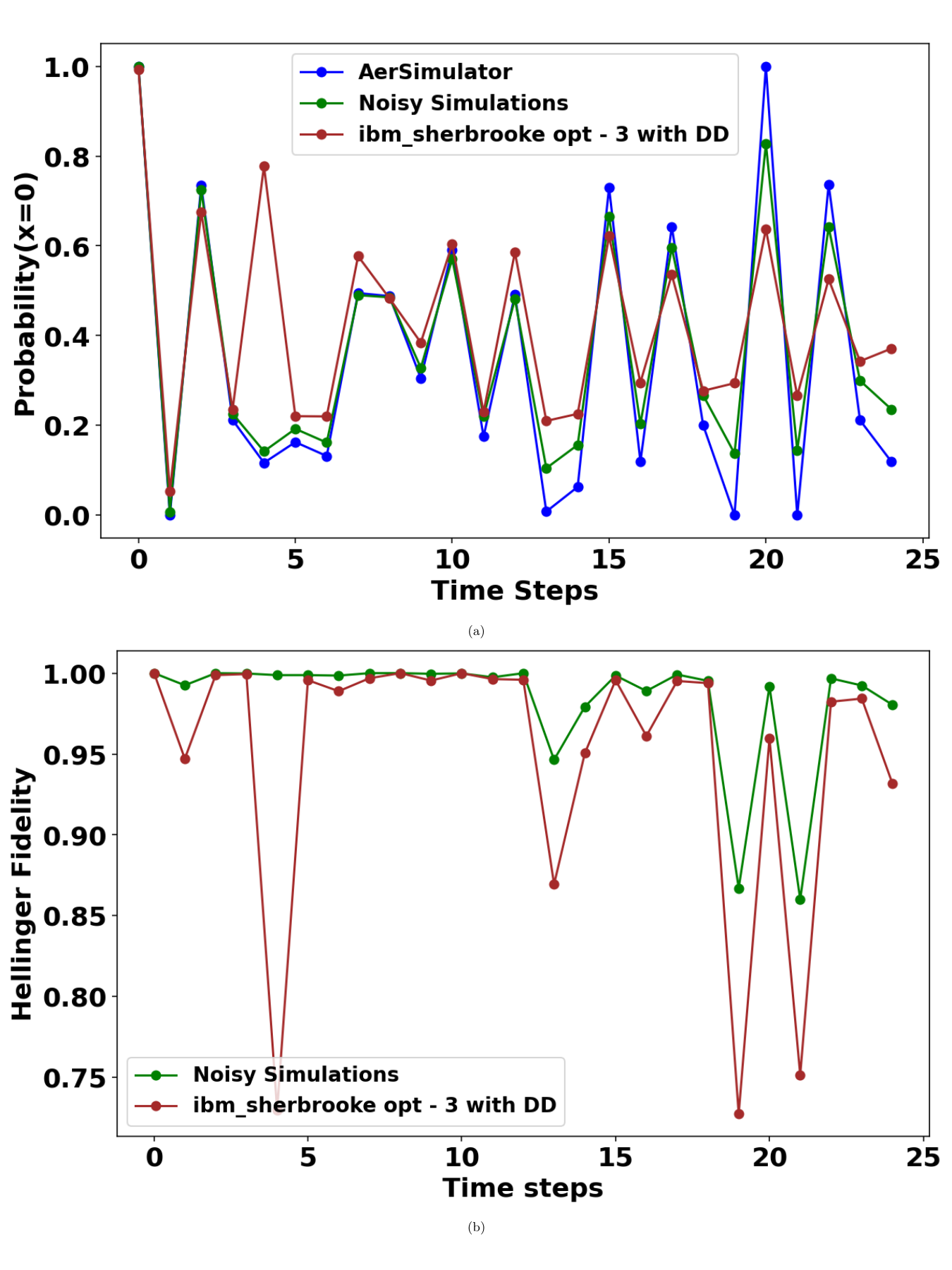}
    \caption{$(a)$ Ordered QW on 3-cycle by Parrondo sequence $A'A'B'B'...$ with periodicity 20 implemented in \texttt{ibm\_sherbrooke} for 25 time steps with DD sequence XY4 with $(b)$ Hellinger fidelity. Results are obtained for $10^5$ shots with position state being encoded in qubits 69,70, and coin state in qubit 74.}
    \label{fig: f19}
\end{figure}

To improve the fidelity in 3-cycle graphs, we apply the $XY4$ DD sequence during the transpilation of the quantum circuit into NISQ hardware. Figs.~\ref{fig: A_dd},~\ref{fig: B_dd}, and ~\ref{fig: f19} show the probability distributions of chaotic and ordered QWs on the 3-cycle graph with improved fidelities due to DD. The NISQ implementation of the DTQW on $3$-cycle shows a signature of emergence of an ordered walk $A'A'B'B'..$ from the combination of chaotic walks due to coins $A'$ and $B'$ with fidelity $>90\%$ in all three NISQ devices up to $25$ time steps. The Hellinger fidelities through the integration of DD improve to $>80\%$ for $A'A'A'..$, $>75\%$ $B'B'B'..$, and $>75\%$ for $A'A'B'B'..$ while maintaining fidelity $~95\%$ at the periodic step, that is $t = 20$ respectively. The saturation of the probability distribution as observed in Figs.~\ref{fig: A},~\ref{fig: B}, and~\ref{fig: AABB} is no longer observed here. 

Although the application of dynamical decoupling (DD) enhances the capture of periodicity, it does not lead to a significant improvement in fidelities. This suggests that the dominant source of noise in the circuit arises from noisy gate operations rather than decoherence during idle periods. Additionally, the circuit depth remains unchanged, indicating that DD only provides marginally better implementation for 3-cycle quantum walks on NISQ devices. To improve the circuit implementation, other error mitigation techniques need to be employed in the current NISQ hardware.
\section{Analysis}\label{4}
\begin{figure}[h]
    \centering
    \includegraphics[width=1\linewidth]{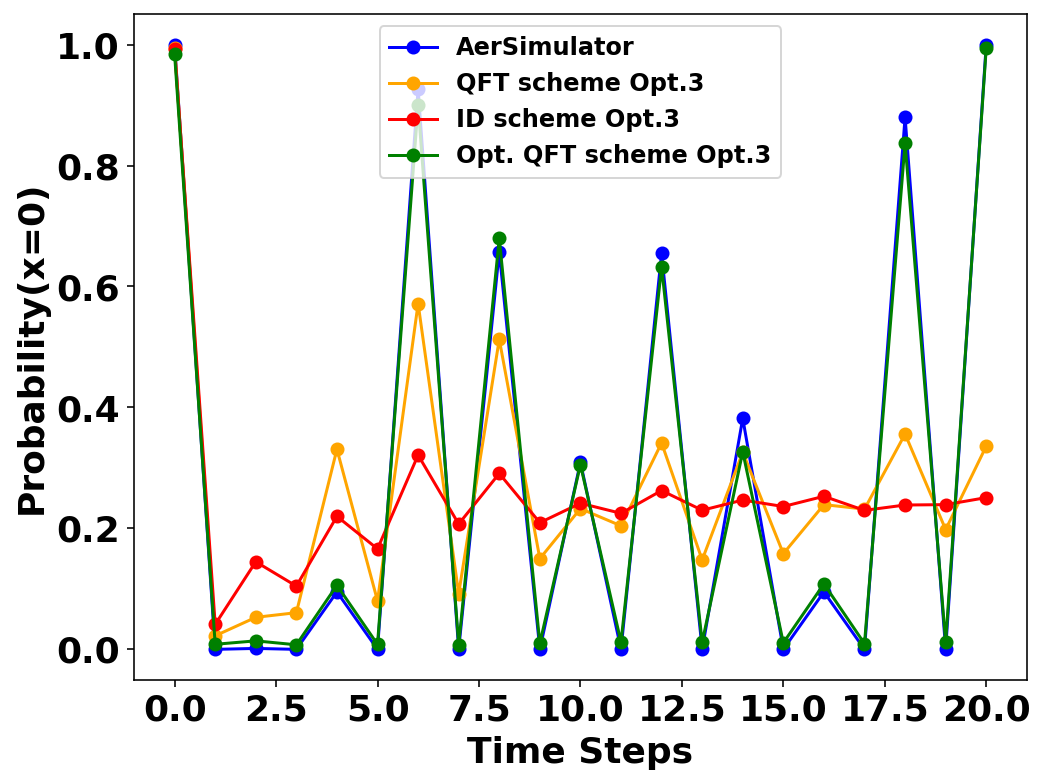}
    \caption{Probability distribution of ordered QW with Parrondo strategy $AABB..$ in 4-cycle implemented with different schemes on \texttt{ibm\_sherbrooke} up to 20 time steps at Opt. level 3 with position state being encoded in qubits 46,47, and coin state in qubit 48.}
    \label{fig:compare}
\end{figure}

As discussed in Section~\ref{2}, there are several circuit proposals to implement DTQW on cyclic graphs on NISQ quantum devices~\cite{DW,SK,Luca}. The methods are based on an increment-decrement scheme involving generalized multi-qubit $CNOT$ gates~\cite{DW}, a QFT-based approach for shift diagonalization~\cite{SK}, and an optimized QFT-based method~\cite{Luca} with minimum resources. However, these methods are limited to even cyclic graphs. The present work utilizes the optimized QFT-based approach and realizes the transition from chaos to order, that is, controlling chaos in three NISQ devices. Using the different available schemes for circuit implementations of DTQWs on even (4) cyclic graphs, we implement our Parrondo strategies $AABB...$ of chaotic coin operators up to 20 time steps and analyze their results in Table~\ref{t2}. The controlled increment-decrement-based scheme~\cite{DW} involves the multi-qubit $CNOT$ gates, which account for the shift of the walker in a cyclic graph at every time step. Since multi-qubit gates cannot be executed in parallel, the depth of the circuit is $O(n^2t-14nt)$~\cite{Luca}. When implemented with the highest level of optimization on an NISQ device, the circuit remains still deep, gathers more noise due to multi-qubit $CNOT$ gates, and does not capture the periodicity of the walk. This scheme is further improved by the QFT-based method~\cite{SK}, which diagonalizes the shift operator by the quantum Fourier transform, thereby reducing the number of multi-qubit gates to simple one- and two-qubit gates. Although this scheme minimizes the depth of the circuit to $O(6nt)$~\cite{Luca}, the implementation of QFT at every time step remains a setback for achieving ordered QW on an NISQ device. Ref.~\cite{Luca} optimizes the QFT-based scheme by exploiting the unitary nature of QFT and significantly reduces the depth of the circuit to $O(nt)$. However, when implemented at the lowest level of optimization (level 1), the circuit remains deep (see, Fig~\ref{fig: dd plot}), but it reduces to a constant one at the highest level of optimization (level 3). The present work implements the optimized QFT approach to realize the transition from chaos to order on three NISQ devices and successfully captures the periodic as well as the chaotic nature of the QW with $98\%$ fidelity for even (4) cyclic graphs. Fig.~\ref{fig:compare} shows the probability of the walker at the initial position ($|000\rangle$) implemented using the above-mentioned schemes realized on \texttt{ibm\_sherbrooke} at optimization level 3. \par
 We further extend our study to odd cycles by modifying the QFT matrix for a 2-qubit system. Although the proposed circuits perform well when simulated in the Qiskit framework (AerSimulator), they do not translate well to the NISQ hardware. This discrepancy is due to the larger circuit depth of $O((n+1)t+2(3n+1))$ (see, Appendix~\ref{depth}), even at the highest level of optimization, and the modified QFT leading to a position disorder in the system~\cite{Dinesh}. This highlights the limitations of current NISQ devices in executing deeper circuits for odd-cycle graphs, despite their correctness in noiseless classical simulations. In order to improve the fidelity values, we integrate our 3-cycle circuits with $XY4$ DD pulses during the transpilation. With the integration of DD-sequences, although the depth does not reduce, the circuits no longer exhibit saturation in probability and show improved fidelity, validating the ordered dynamics on NISQ hardware. Table~\ref{t3} analyzes the depth and fidelity values with and without DD for the 3-cycle when implemented on an NISQ device.
\begin{table}[h]
    \centering
    \renewcommand{\arraystretch}{1.1}
    
    \begin{tabular}{|l|c|c|c|}
        \hline
        \textbf{Scheme} &\textbf{Depth} & \textbf{Period captured} &\textbf{Fidelity}  \\
        \hline
        {Cont. Incr.-Decr.~\cite{DW}} & {600} & No & 70$\%$\\
        \hline
        {QFT~\cite{SK}} & {585} & No & ~80$\%$\\
        \hline
        {Opt.QFT~\cite{Luca} \ref{3a}}& {5} & Yes & $98\%$\\
        \hline
    \end{tabular}
    \caption{Controlling chaos via Parrondo strategy $AABB..$ using different circuit schemes for 4-cycle DTQWs implemented in \texttt{ibm\_sherbrooke} at Opt. level 3 at $t=20 $ time step with position state being encoded in qubits 46,47, and coin state in qubit 48.}
    \label{t2}
\end{table}%
 \begin{table}[h]
    \centering
    \renewcommand{\arraystretch}{1.1}
    
    \begin{tabular}{|l|c|c|c|c|}
        \hline
        \textbf{Scheme}  & \textbf{Depth} & \textbf{\shortstack{Period \\captured}}&\textbf{\shortstack{Fidelity \\at $t=20$}}&\textbf{\shortstack{Fidelity\\(25 steps)} }\\
        \hline
        {\shortstack{Opt. QFT \\without DD~\ref{3c}}} & {515} & No &~90$\%$ &$>65\%$\\
        \hline
        {\shortstack{Opt. QFT \\with DD~\ref{DD}}} & {515} &Yes& $~95\%$&$>75\%$\\
        \hline
    \end{tabular}
    \caption{Controlling chaos via Parrondo Strategy $A'A'B'B'..$ using different schemes for 3-cycle DTQWs implemented in \texttt{ibm\_sherbrooke} with and without DD sequence $XY4$ at Opt. level 3 up to 25 time steps with position state being encoded in qubits 69,70, and coin state in qubit 74.}
    \label{t3}
\end{table}%
\section{Conclusion}\label{con}
This work presents a study on the visualization of the transition from chaos to order on NISQ devices via cyclic graphs. We established a general scheme for quantum circuit implementations of DTQW on both even- and odd-cycle graphs using the optimized QFT-based approach modified via dynamical decoupling for odd cycles. We further propose a quantum circuit to realize the DTQW in three different NISQ quantum computers. Our work focused on visualizing the transition of chaos to order in real quantum hardware (provider: IBM). Applying Parrondo's strategy, we demonstrate the emergence of order from chaos in both even- and odd-cycle graphs (3 and 4-cycle) in three NISQ quantum computers. Our results for the 4-cycle using the optimized QFT scheme are in good agreement with the theoretical predictions (fidelity $>98\%$); however, due to the large depth and inherent noise in the hardware, the results of the 3-cycle circuit (fidelity $65$-$90\%$) did not agree well with the ideal simulation produced by AerSimulator. The integration of dynamical decoupling pulses into odd-cycle circuits resulted in an improvement in fidelity ($75$-$95\%$). Our scheme provides a general idea of circuit implementations of arbitrary cyclic DTQW, hence opening avenues for various practical applications. In parallel, we are executing a cryptographic protocol based on periodic QWs that builds on the scheme introduced here~\cite{ARath}. Future efforts will focus on learning methods to reduce the depth of quantum circuits in odd-cycle graphs by employing several other error mitigation techniques to enhance their performance. This work will significantly contribute to the state-of-the-art NISQ implementations of entanglement-based cryptographic protocols to study the generation of maximally entangled single-particle states in real quantum hardware.

\onecolumngrid
\appendix
\hspace{6.6cm}{\textbf{\large APPENDIX}}
\vspace{0.5cm} 

Here, we describe additional details that support the findings of the main text. In Appendix~\ref{Algorithm}, we provide the algorithms used to visualize the transition from chaos to order in even (4) and odd (3) cycle graphs using Parrondo strategies. Appendix~\ref{trans} presents the transpiled quantum circuits of the 3 and 4-cycle graphs for the NISQ devices for one time step of the QW. Appendix~\ref{depth} includes a calculation of the circuit depth for the 3-cycle implementation, based on modified quantum Fourier transforms (QFTs), and provides a general expression for depth as a function of time steps. Appendix~\ref{results} reports results from real hardware experiments conducted on \texttt{ibm\_brisbane} and \texttt{ibm\_kyiv}, including fidelity and probability distributions for 4- and 3-cycle walks, with and without DD techniques. Appendix~\ref{5} describes the calculation of Lyapunov exponent, which confirms whether we have chaotic and periodic quantum walks on cyclic graphs. Appendix~\ref{6} shows the probability distributions of quantum walks generated from combining the chaotic unitary operators $A$ and $B$ in sequences, $ABA...$, $ABB..$, and $AAB...$ on 4-cycle graphs. Appendix~\ref{7} provides the results obtained from comparing the probability distributions using Hellinger fidelity and Bhattacharya fidelity. 
\section{Algorithm for the transition of chaos to order via cyclic DTQWs}\label{Algorithm}[H]
Here, we present the algorithms for the circuit implementation of DTQWs to realize the transition from chaos to order in even(Fig.~\ref{fig:quantum_circuit}) and odd(Fig.~\ref{fig:quantum_circuit3}) cycle graphs.
\begin{algorithm}[H]
\caption{4-Cycle DTQW with Parrondo Strategy AABB..}
\label{alg:hamiltonian_path}
\begin{algorithmic}[1]
\State \textbf{Input:} Number of time steps $t$, Chaotic coin matrices $A$ and $B$
\State Create a quantum circuit with 3 qubits: $q_2$ (coin), $q_1q_0$ (position)
\State Apply QFT to position qubits $q_1q_0$ \Comment{Move to Fourier basis}
\For{$i = 0$ to $t - 1$}
    \If{$i \bmod 4 = 0$ or $i \bmod 4 = 1$} \Comment{Coin Operator}
        \State Apply coin operator $A$ to $q_2$
    \Else
        \State Apply coin operator $B$ to $q_2$
    \EndIf
    \State Apply phase gate $P(-\pi)$ to $q_0$  \Comment{Shift Operator}
    \State Apply phase gate $P(-\pi / 2)$ to $q_1$
    \If{$q_2$ is in state $\ket{1}$}
        \State Apply controlled phase gate $P(\pi)$ to $q_1$
    \EndIf
\EndFor
\State Apply inverse QFT to position qubits $q_1q_0$ \Comment{Back to position basis}
\State Measure the position qubits $q_1q_0$
\end{algorithmic}
\end{algorithm}
\begin{algorithm}[H]
\caption{3-Cycle DTQW with Parrondo Strategy A'A'B'B'..}

\begin{algorithmic}[1]
\State \textbf{Input:} Number of time steps $t$, Chaotic coin matrices $A'$ and $B'$
\State Create a quantum circuit with 3 qubits: $q_2$ (coin), $q_1q_0$ (position)
\State Apply QFT to position qubits $q_1q_0$ \Comment{Move to Fourier basis}
\For{$i = 0$ to $t - 1$}
    \If{$i \bmod 4 = 0$ or $i \bmod 4 = 1$} \Comment{Coin Operator}
        \State Apply coin operator $A'$ to $q_2$ 
    \Else
        \State Apply coin operator $B'$ to $q_2$
    \EndIf
    \State Apply phase gate $P(-4\pi/3)$ to $q_0$ \Comment{Shift Operator}
    \State Apply phase gate $P(-2\pi / 3)$ to $q_1$
    \If{$q_2$ is in state $\ket{1}$}
        \State Apply controlled phase gate $P(4\pi/3)$ to $q_1$
         \State Apply controlled phase gate $P(8\pi/3)$ to $q_0$
    \EndIf
\EndFor
\State Apply inverse QFT to position qubits $q_1q_0$ \Comment{Back to position basis}
\State Measure the $q_1q_0$ position qubits.
\end{algorithmic}
\end{algorithm}
The codes that support the plots of the main text are available in Ref.~\cite{github}.
\section{Transpiled quantum circuits in NISQ hardware}
\label{trans}
\begin{figure}[H]
    \centering
    \includegraphics[width=1\linewidth]{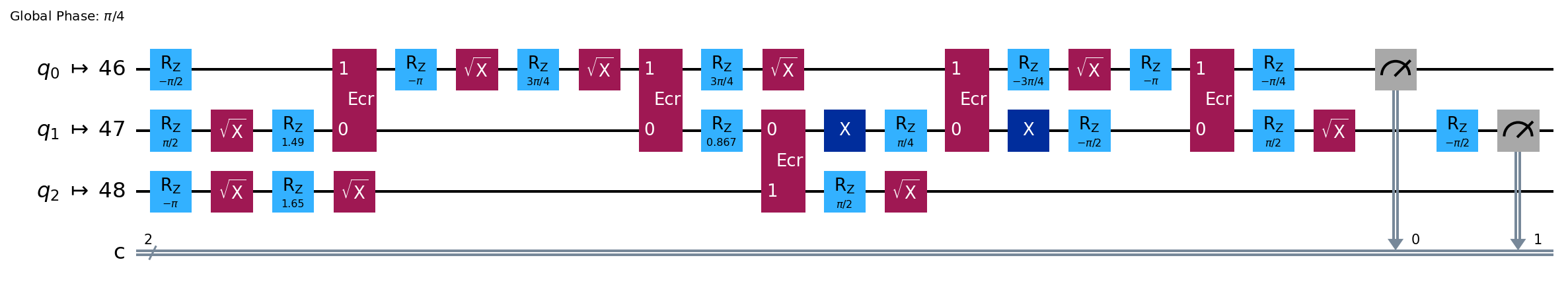}
    \caption{Quantum circuit implementing DTQW on 4-cycle with unitary operator $A$, transpiled in \texttt{ibm\_sherbrooke} at optimization level 3 for $t=1$ step.}
    \label{s1}
\end{figure}

\begin{figure}
    \centering
    \includegraphics[width=1\linewidth]{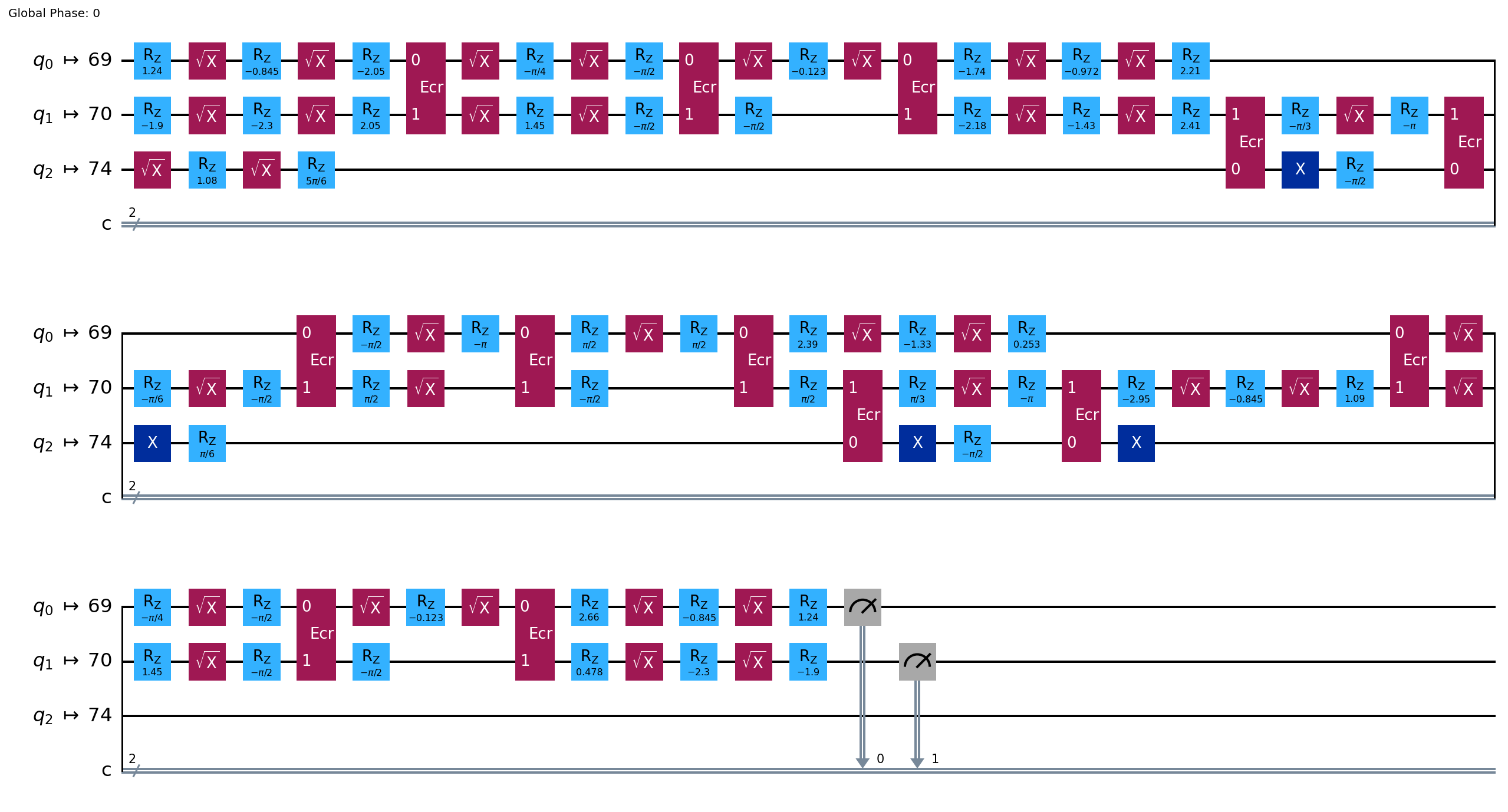}
    \caption{Quantum circuit implementing DTQW on 3-cycle with unitary operator $A$, realized in \texttt{ibm\_sherbrooke} at optimization level 3 for $t=1$ step.}
    \label{s2}
\end{figure}

The quantum circuits described in Sec.~\ref{3} are designed within the Qiskit framework. When these circuits are fed to a quantum computer, they are transpiled by the processor, that is, the abstract circuit designed is converted into the native gates depending upon the level of optimization. The quantum computers that we consider for our study are \texttt{ibm\_sherbrooke}, \texttt{ibm\_brisbane}, and \texttt{ibm\_kyiv} which are three quantum computers under the \textit{Eagle r3} processor whose native gates~\cite{ibm} are, \\
\textbf{(a)} Identity Gate ID = $\begin{pmatrix} 1 & 0 \\ 0 & 1 \end{pmatrix},$\\  
\textbf{(b)} $R_z(\theta)$ = $\begin{pmatrix} e^{-i\theta /2} & 0 \\ 0 & e^{i\theta /2} \end{pmatrix}$ is a single qubit rotation of angle $\theta$ about the Z-axis,\\
\textbf{(c)} X = $\begin{pmatrix} 0 & 1 \\ 1 & 0 \end{pmatrix},$  is the single qubit Pauli-X gate which is equivalent to a rotation of angle $\pi$ about the X-axis (global phase of $\pi/2$), \\
\textbf{(d)} SX = $\sqrt{X} = \frac{1}{2} \begin{pmatrix} 1+i & 1-i \\ 1-i & 1+i \end{pmatrix}$ is a single qubit gate that implements a rotation of angle $\pi/2$ about the X-axis with a global phase of $\pi/4$, and\\
\textbf{(e)} Echoed Cross-Resonance (ECR) = $\frac{1}{\sqrt{2}} \begin{pmatrix} 0 & 1 & 0 & i \\ 1 & 0 & -i & 0 \\ 0 & i & 0 & 1 \\ -i & 0 & 1 & 0 \end{pmatrix} = \frac{1}{\sqrt{2}}({ID\otimes X - X \otimes Y})$ is a two-qubit maximally entangling gate equivalent to CNOT up to single-qubit pre-rotations where X and Y are Pauli-X and Y gates. Every gate of a quantum circuit is converted into terms of the native gates of a processor by the transpiler. For example, the chaotic coin operator $A = C_2(r = 0.998489,a = 0, b=0)$ for 4-cycle DTQW, is transpiled into, $\sqrt{X}R_Z(3.06)\sqrt{X}$ and  the chaotic coin operator $A' = C_2(r = 0.264734,a = 0, b=0)$ for 3-cycle DTQW, is transpiled as $\sqrt{X}R_Z(1.08)\sqrt{X}$.\\
 Figs.~\ref{s1} and \ref{s2} show the transpiled circuits implementing 1 time step at optimization level 3 for 4 and 3-cycles realized in \texttt{ibm\_sherbrooke}, respectively. As discussed in Sec.~\ref{3c}, in order to implement 3-cycle DTQW on an NISQ device, we modify the QFT applied to 4-cycle and add more controlled operations so as to keep the unused sites isolated. This, in turn, increases the size of the quantum circuit even when transpiled at the highest level of optimization.
\section{Calculation of Depth of Proposed 3-cycle quantum circuit}
\label{depth}
\begin{figure}[H]
    \centering
    \includegraphics[width=0.8\linewidth]{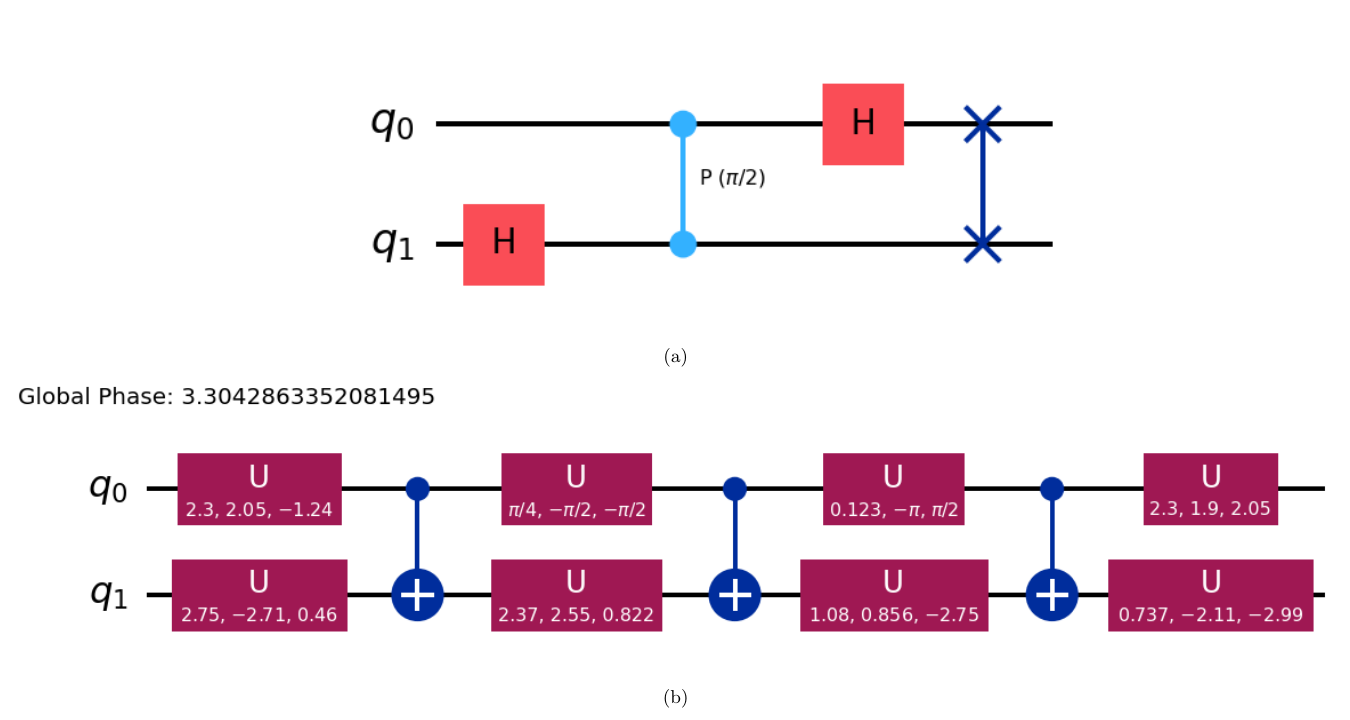}
    \caption{(a) QFT circuit for 4-cycle and (b) modified QFT circuit for 3-cycle where U gates are the general single qubit rotation gates defined in Eq.(\ref{U}).}
    
    \label{fig: qft}
\end{figure}

In order to implement the 3-cycle graph on a quantum circuit, we modify the QFT matrix, as it accounts for the shift of the walker on the cycle, given in Eq.(\ref{e22}) of  Sec.~\ref{3c}. Fig.~\ref{fig: qft} shows the modified QFT circuit for the 3-cycle graph. It is evident that QFT (IQFT) adds $4n$ one-qubit gates, $n+1$ two-qubit gates, and has depth $(3n+1)$. 
  The IQFT circuits have an architecture similar to that of QFT circuits, but are applied in the opposite order. The $U$ gates are general single-qubit gates with three parameters, $\theta,\phi,\lambda$~\cite{qiskit}, required for rotations in a quantum circuit having the form, 
\begin{equation}
    U(\theta,\phi,\lambda) = \begin{pmatrix}
        \cos({\theta/2})& -e^{i \lambda/2}\sin({\theta/2})\\e^{i \phi/2}\sin({\theta/2})&e^{i (\lambda+\phi)/2}\cos({\theta/2})
    \end{pmatrix} .
    \label{U}
\end{equation}
From Fig.~\ref{fig:quantum_circuit3} of Sec.~\ref{3c}, we require $n+1$ qubits ($n=2$) to implement the DTQW of the 3-cycle graph on a quantum circuit. We note that the $n$ controlled-phase gates have the same control qubit but different target qubits, and thus, are executed sequentially. This results in adding a depth $n$ at every time step. The coin gate and the layer of $n$ $\tilde{P_i}$ phase gates (not controlled) can be executed in parallel at every time step, hence amount to a unit depth in the quantum circuit. Also, the QFT and IQFT are applied once at the beginning and once at the end, respectively, each accounting for a depth of $3n+1$ (from Fig.~\ref{fig: qft}(b)). Thus, in total: 
%From Fig.~\ref{fig:quantum_circuit3} of Sec.~\ref{3c}, we note that the controlled-phase gates are applied sequentially, hence adding a depth of $n$ at each time step. The coin gate and the layer of $\tilde{P_i}$ phase gates (not controlled) can be executed in parallel, hence amount to a unit depth in the quantum circuit. Also, the QFT and IQFT are applied at the beginning and at the end, respectively, each accounting for a depth of $3n+1$. The details of the quantum circuit are listed below.  
%The $P_k^2$ gates add a depth of $n$ at each time step as they are applied sequentially; the coin and the layer of $R_k^\dagger$ gates amount to unit depth as they are executed in parallel.
\begin{itemize}
    \item[1.] No. of one-qubit gates = $2(4n) + t(1 + n )$ ($4n$ due to QFT(IQFT), $n$ $\tilde{P_i}$ gates and 1 coin)
    \item [2.] No. of two-qubit gates = $2(n+1) + nt$ ($n+1$ due to QFT(IQFT) and $n$ for controlled $\tilde{P_i}$ gates)
    \item [3.] Depth = $2(3n+1) + t(1 + n) $
\end{itemize}
\section{Results from \texttt{ibm\_brisbane} and \texttt{ibm\_kyiv}}
\label{results}
\subsection{NISQ implementation of Parrondo Strategy \texorpdfstring{$AABB..$}{AABB..} on 4-cycle graphs}
\label{results1}
Here we provide the results of the transition from chaos to order in even cyclic graphs (4-cycle) realized in NISQ quantum computers, \texttt{ibm\_brisbane} and \texttt{ibm\_kyiv} at optimization level 3. Both devices capture the periodicity of the QW, that is, 20, with a fidelity $>95\%$ throughout the 25 time steps. However, \texttt{ibm\_kyiv} performs better than \texttt{ibm\_brisbane} by maintaining fidelity of $>98\%$ up to 25 time steps. The circuit depth of both devices is the same as that of \texttt{ibm\_sherbrooke} as described in Section~\ref{4}. Table~\ref{t2_3} compares the different NISQ devices according to the encoded qubits, the depth of the quantum circuit, and the fidelity values. The transpiler chooses the set of qubits that provides the highest fidelity based on the level of optimization. Although all three devices are transpiled at the highest level of optimization (level 3), \texttt{ibm\_sherbrooke} and \texttt{ibm\_kyiv} perform slightly better than \texttt{ibm\_brisbane}.
\begin{figure}[H]
    \centering
    \includegraphics[width=0.9\linewidth]{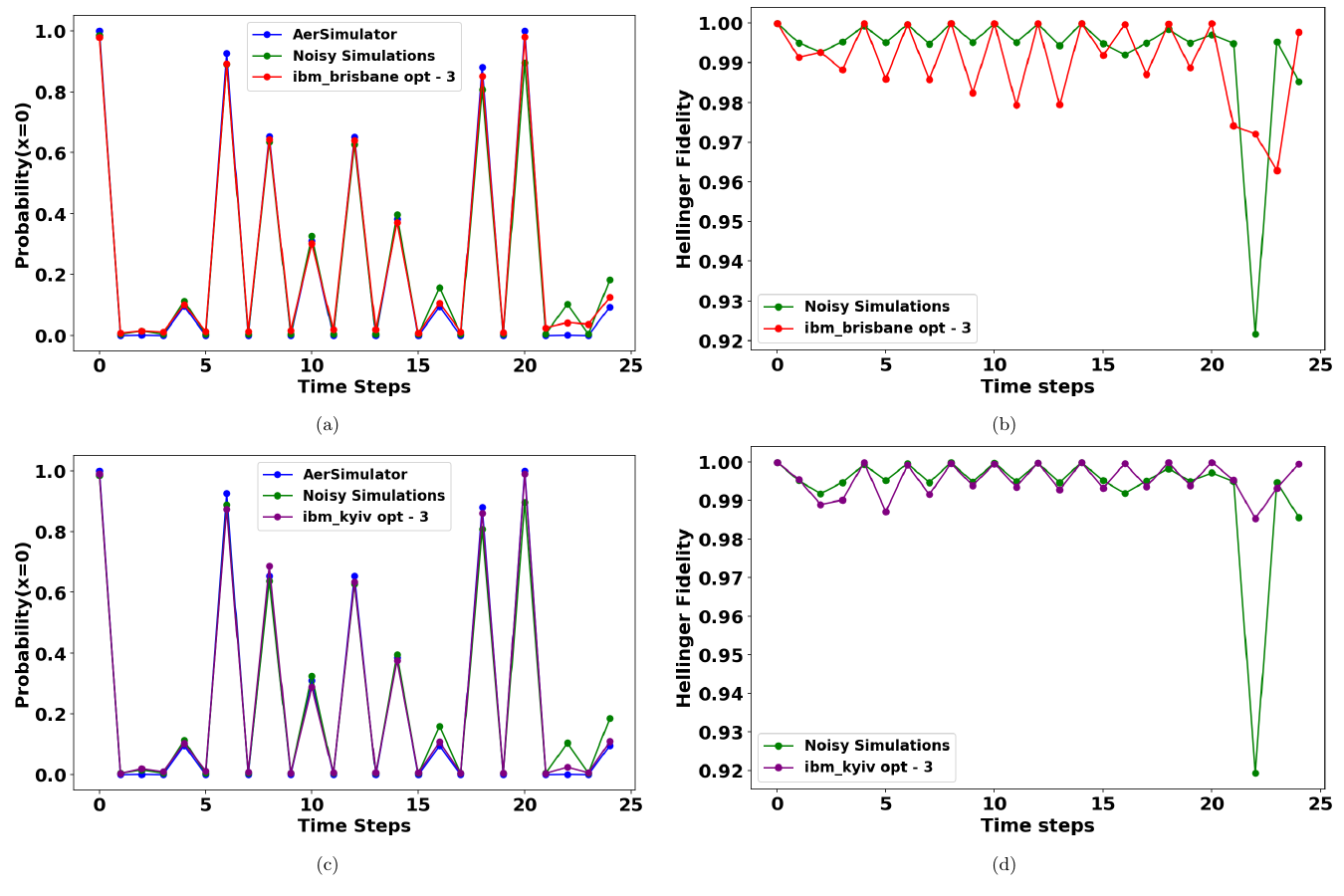}
    \caption{Probability distribution of the walker at initial site and Hellinger fidelity vs time steps with circuits($AABB..$) implemented in (a),(b)\texttt{ibm\_brisbane} and (c),(d)\texttt{ibm\_kyiv} on $4$-cycle. Results are obtained for $10^5$ shots with position state being encoded in qubits 3, 4, and coin state in qubit 5 with \texttt{ibm\_brisbane} and those respectively in qubits 41, 53, and 60 with \texttt{ibm\_kyiv}.}
    \label{fig: AABB probs}
\end{figure}
\begin{table}[h]
    \centering
    \renewcommand{\arraystretch}{1.1}
    
    \begin{tabular}{|l|c|c|c|}
        \hline
        \textbf{NISQ device} &\textbf{\shortstack{Qubits encoded\\($q_0$-$q_1$-$q_2$)}}&\textbf{Depth} &\textbf{Fidelity} \\
        \hline
        {\texttt{{ibm\_sherbrooke}}} & 46-47-48 & {5} &$>98\%$\\
        \hline
        {\texttt{{ibm\_brisbane}}} & 3-4-5& {5}  & $\sim 95\%$\\
        \hline
        {\texttt{{ibm\_kyiv}}} & 41-53-60& {5} & $>98\%$\\
        \hline
    \end{tabular}
    \caption{Comparison of NISQ devices for visualizing transition from chaos to order for 4-cycle DTQW implemented at optimization level 3 at time step $t=20$.}
    \label{t2_3}
\end{table}%
\subsection{NISQ implementation of Parrondo Strategy \texorpdfstring{$A'A'B'B'..$}{A'A'B'B'..} on 3-cycle graphs}
\label{results2}
Here we provide the results of the transition from chaos to order in odd cyclic graphs (3-cycle) realized in NISQ quantum computers, \texttt{ibm\_brisbane} and \texttt{ibm\_kyiv} at optimization level 3 integrated with DD sequence, $XY4$. Both devices capture the periodicity of the QW, that is, 20, with a fidelity $>90\%$ throughout the 25 time steps. Table~\ref{t2_4} compares the different NISQ devices according to the encoded qubits, the depth of the quantum circuit, and the fidelity values. The transpiler chooses the set of qubits that provides the highest fidelity based on the level of optimization. Although all three devices are transpiled at the highest level of optimization (level 3), \texttt{ibm\_sherbrooke} performs better than \texttt{ibm\_brisbane} and \texttt{ibm\_kyiv}.
\begin{figure}[H]
    \centering
    \includegraphics[width=1\linewidth]{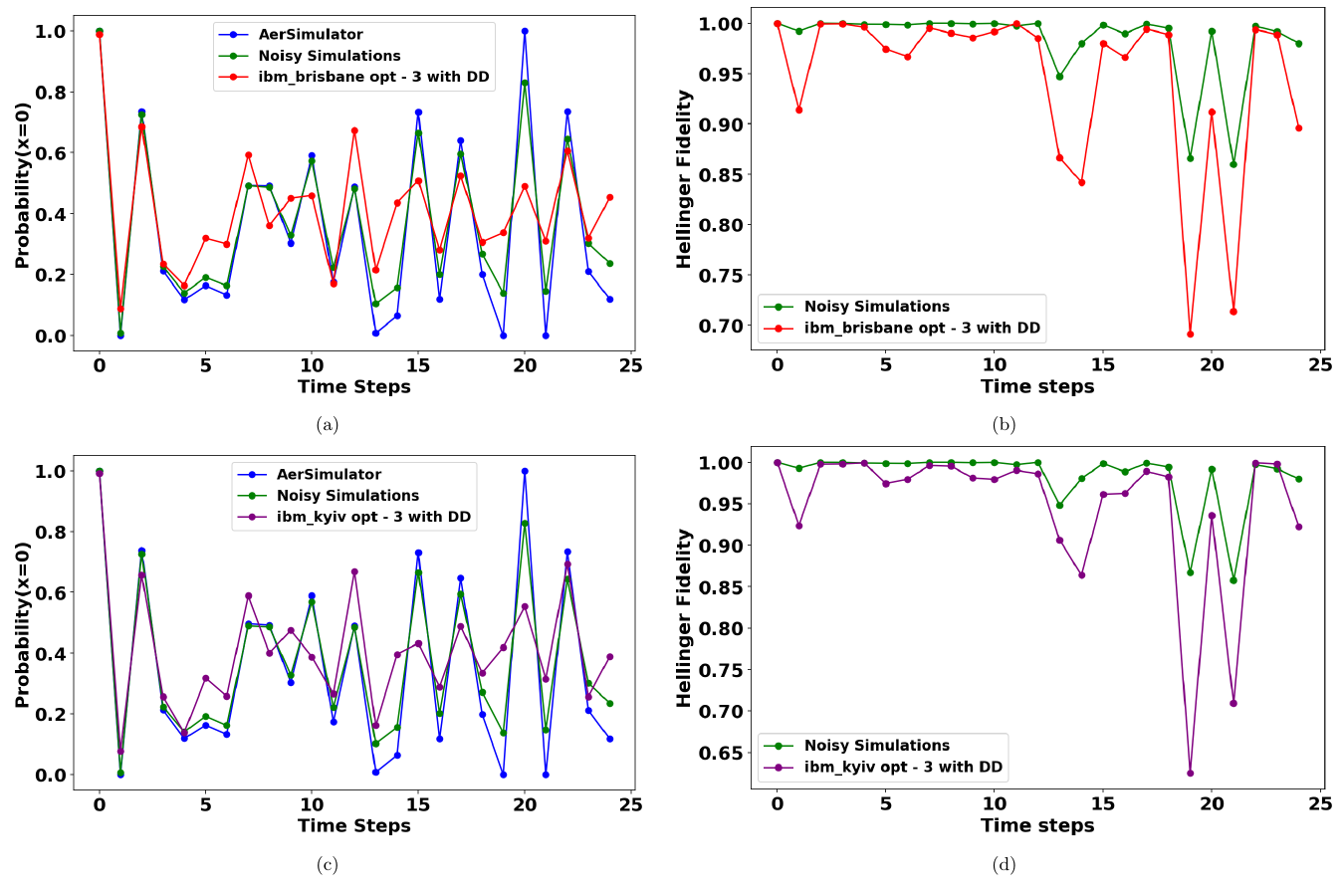}
    \caption{Probability distribution of the walker at initial site and Hellinger fidelity vs time steps with circuits($AABB..$) implemented in \texttt{ibm\_brisbane} and \texttt{ibm\_kyiv} for $3$-cycle with DD sequence $XY4$. Results are obtained for $10^5$ shots with position state being encoded in qubits 31, 32, and coin state in qubit 36 with \texttt{ibm\_brisbane} and those respectively in qubits 69, 70, and 74 with \texttt{ibm\_kyiv}.}
    \label{fig: AABB_3 probs}
\end{figure}

\begin{table}[H]
    \centering
    \renewcommand{\arraystretch}{1.1}
    
    \begin{tabular}{|l|c|c|c|c|}
        \hline
        \textbf{NISQ device} &\textbf{\shortstack{Qubits encoded\\($q_0$-$q_1$-$q_2$)}}&\textbf{Depth} &\textbf{\shortstack{Fidelity\\at t =20}} &\textbf{\shortstack{Fidelity overall \\(up to 25 steps)}} \\
        \hline
        {\texttt{{ibm\_sherbrooke}}} & 69-70-74 & {515} &$\sim95\%$ &$>75\%$\\
        \hline
        {\texttt{{ibm\_brisbane}}} & 31-32-36& {520}  & $\sim 90\%$ &$>70\%$\\
        \hline
        {\texttt{{ibm\_kyiv}}} & 69-70-74& {520} & $\sim90\%$ &$>65\%$\\
        \hline
    \end{tabular}
    \caption{Comparison of NISQ devices for visualizing transition from chaos to order for 3-cycle DTQW implemented at optimization level 3 at up to time steps using $XY4$ DD sequence.}
    \label{t2_4}
\end{table}%

\section{Calculation of Lyapunov Exponent}\label{5}
In sections~\ref{2} and \ref{3}, we use the same definition of chaotic quantum walks as stated in Ref.~\cite{treganna}, that is, if Eq.(\ref{e8}) is satisfied, the quantum walk is periodic (ordered, i.e., walker returns to initial position with probability 1), else it is chaotic. However, Ref~\cite{panda,lyapunov} also provides another way of confirming whether cyclic discrete-time quantum walks are chaotic or not, based on the Lyapunov exponent~\cite{lyapunov}. A positive value of Lyapunov exponent $\alpha$ indicates a chaotic walk, while $\alpha=0$ indicates an ordered quantum walk. We briefly outline the recipe for calculating the Lyapunov exponent in cyclic quantum walks, and then determine it explicitly for the case of a 4-cycle walk.
To calculate the Lyapunov coefficient, we start with an initial
 normalized state at time $t = 0$ on a cyclic graph with $4$ sites, $\ket{\psi(0)}$. Then we let the initial state evolve with, \begin{equation}
 \ket{\psi(t)} = U^t\ket{\psi(0)},
 \label{le}
 \end{equation} with $t>0$. We now end up with two cases, one where the unitary operator yields a chaotic walk, $U_C$, and the other where it yields an ordered walk, $U_p$, with period $T$.
 Thus, \begin{equation}
     \ket{\psi(t=T)} = U_p^T\ket{\psi(0)}=\ket{\psi(0)},\label{le1}\end{equation} that is, $U_p^T=I$, where $I$ is the identity matrix. On the other hand, \begin{equation}
          \ket{\psi(t)} = U_c^t\ket{\psi(0)}\ne\ket{\psi(0)}, \text{ }\text{for any}\text{ }t .\label{le2}
     \end{equation}
Thus, one can define a ``distance'' state, i.e., $\left|\psi_{d}\right\rangle=\left|\psi\left(t\right)\right\rangle-\left|\psi\left(0\right)\right\rangle$ with, $|\psi_{d}\rangle=0$ for $U=U_{p}$ at $t=T$ while $|\psi_{d}\rangle\neq 0$ for $U=U_{c}, \text{ for any } t.$ This probability distance function $d(t)$ can similarly, to Ref.~\cite{lyapunov}, help us calculate the Lyapunov exponent.  The probability distance function can be expanded as
\begin{equation}
\begin{split}
d(t)&=|\langle\psi_{d}\mid\psi_{d}\rangle|=(\langle\psi(t)|-\langle\psi(0)|) (\left|\psi\left(t\right)\right\rangle-\left|\psi\left(0\right)\right\rangle),\\&= |2-2\langle\psi(t)\mid\psi(0)\rangle|=f(\alpha,t).
\end{split}
\label{eE3}
\end{equation}
In the above equation, $\alpha$ is the Lyapunov exponent. The distance function $f(\alpha,t)$ is bounded by maximum value $2$ and the minimum $0$. In chaotic case, $\alpha > 0$ and maximum possible value is $2$, while for periodic case at $t=T$, $\alpha=0$ and $d=0$, this implies $f(\alpha,t)=2(1-2^{-\alpha t})$ and for the Lyapunov exponent one obtains $\alpha$, \begin{equation}
         \alpha=-\frac{1}{t}\log_{2}|\langle\psi(t)\mid\psi(0))\rangle .\label{le3}
\end{equation} Following this for 4-cycle QW, we obtain the positive Lyapunov exponents for chaotic walks $AAAA...$ and $BBBB...$ for any value of $t$, for example, at $t=20$, $\alpha$ for $AAAA..$ and $BBBB..$ is 0.012. We also verified that $\alpha$ remains 0 for the quantum walk with unitary operator sequence, $AABB...$ for time steps $T = 20,40,60...$ Thus, we find that the Lyapunov exponent takes positive values for chaotic quantum walks on the 4-cycle graph, whereas it vanishes for the periodic case. One can find similar results for the 3-cycle graph listed in Ref.~\cite{panda}.

\section{Chaotic Parrondo's Sequences}\label{6}
In the main text, we have focused exclusively on the $AABB…$ sequence of unitary evolution operators. This choice is deliberate: as shown in Ref.~\cite{panda}, only a restricted set of deterministic alternations between the two operators leads to ordered dynamics, with $AABB...$ being the canonical example. Other natural candidates, such as $ABA…$, $ABB…$, or related variants, fail to generate periodic recurrences and remain chaotic. For completeness, we have carried out numerical simulations of these alternative sequences, and the corresponding results are presented below, see Fig.~\ref{23}.

\begin{figure}[H]
    \centering
    \includegraphics[width=1\linewidth]{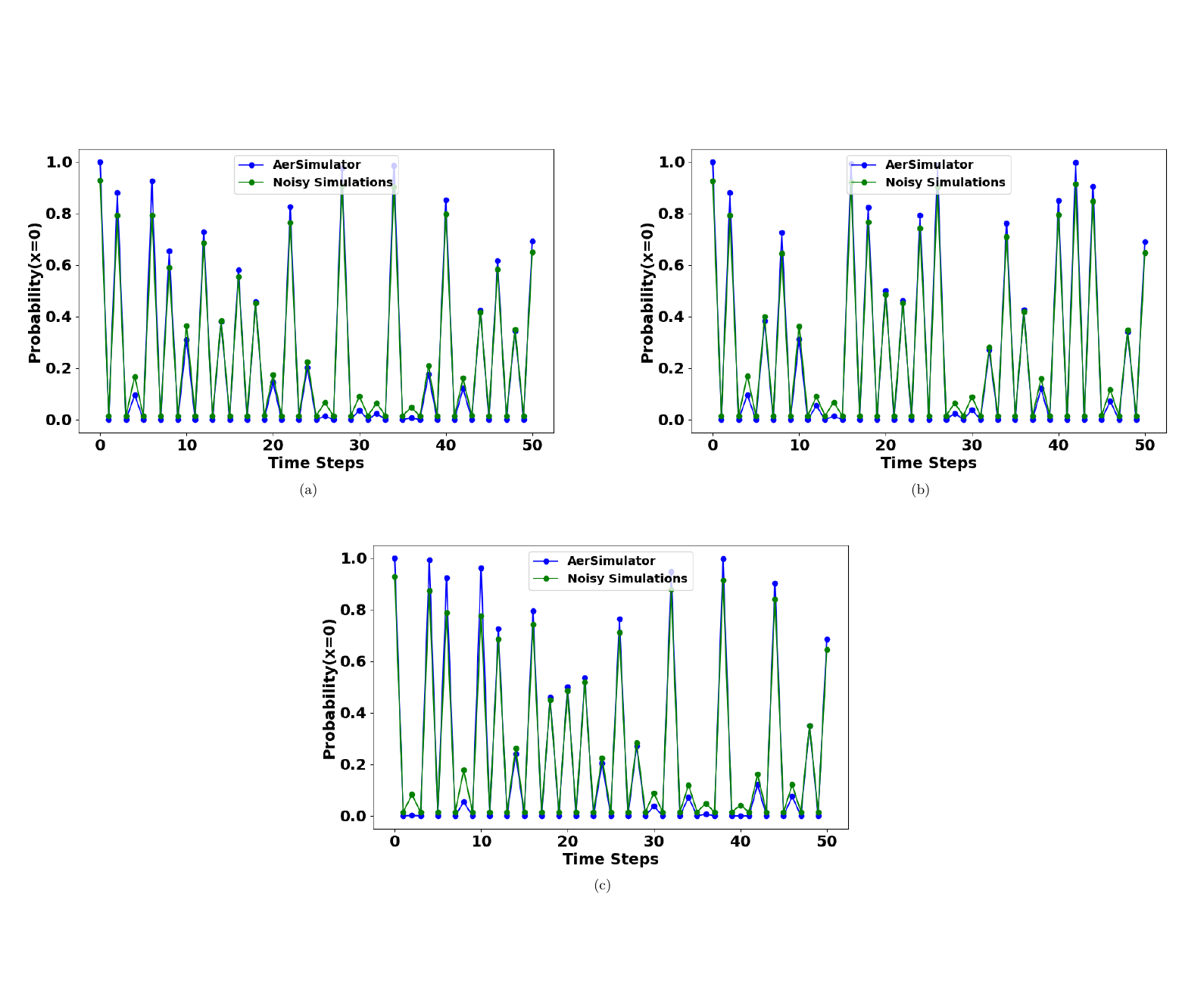}
    \caption{Probability at initial site($x=0$) vs time steps for unitary operator sequences (a)$ABA...$, (b)$ABB..$, and (c)$AAB...$ up to $50$ time steps on 4-cycle graph. All the sequences yield chaotic quantum walks.}
    \label{23}
\end{figure} As expected, the probability dynamics for $ABA...$, $ABB...$, and similar patterns remain chaotic. This justifies our emphasis on the $AABB...$ sequence in the main text, as it provides the demonstration of a chaos-to-order transition in DTQWs. In principle, one may use the method described in Ref.~\cite{Dukes} to obtain various sequences that yield ordered quantum walks.
\section{Results obtained using Bhattacharya fidelity.}\label{7}
Herein, we present the results obtained using Hellinger fidelity~\cite{HF,HF_1} and Bhattacharya fidelity~\cite{BC} as metrics to demonstrate the closeness between the ideal and noisy probability distributions obtained via \texttt{qiskit\_aer.}\begin{figure}[H]
    \centering
    \includegraphics[width=1\linewidth]{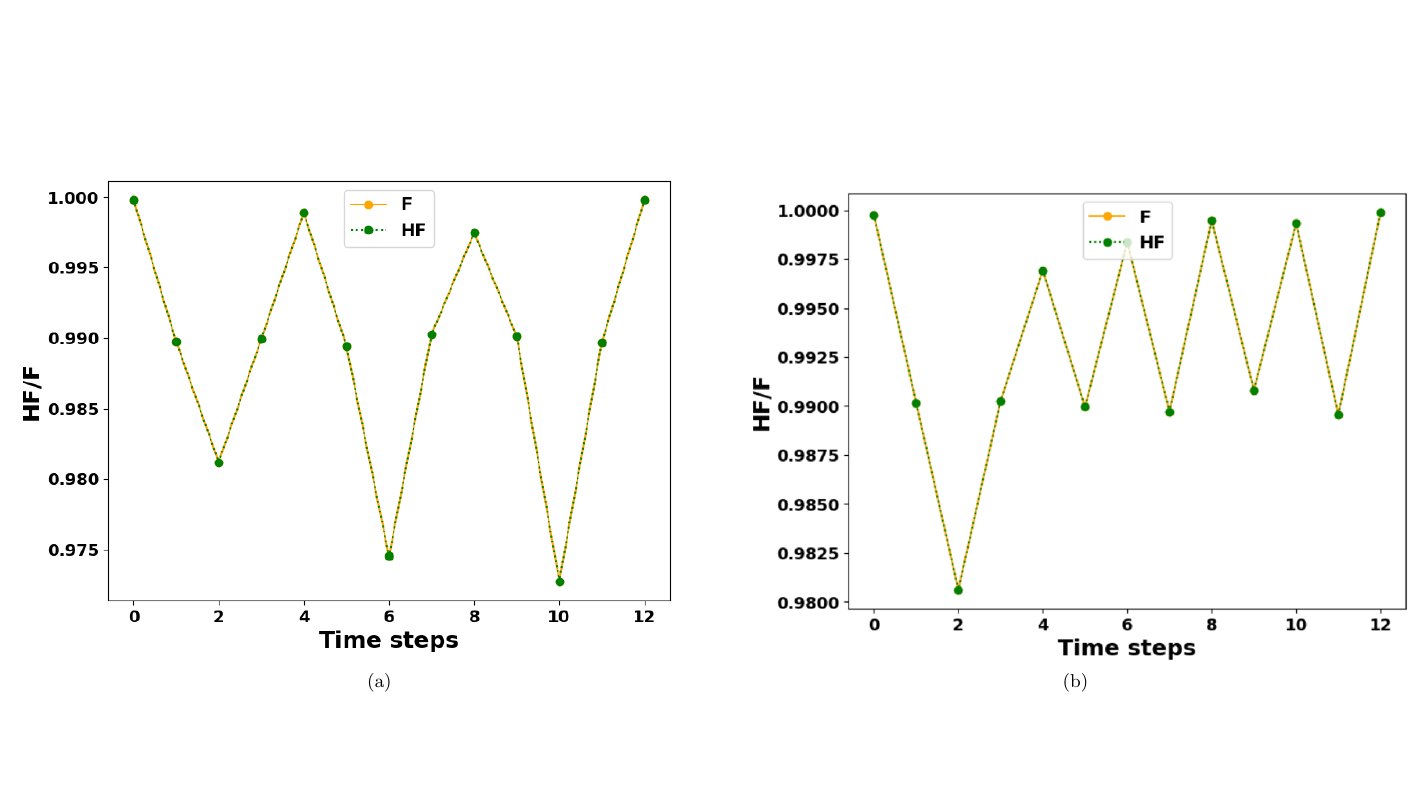}
    \caption{Hellinger fidelity (HF) and Bhattacharya Fidelity (F) vs time steps for unitary coin sequences (a) $AAAA...$ and (b)$AABB...$ up to 12 time steps obtained via \texttt{qiskit\_aer}}
    \label{24} \end{figure}From Fig.~\ref{24}, it is clear that the Hellinger fidelity and Bhattacharya fidelity are equivalent to each other.
\twocolumngrid

\end{document}